%% file: main.tex
\documentclass[letterpaper, 12pt]{article} 
\input{preamble}

\title{Government Transparency and Innovation: \\ Evidence from Wireless Products}
\author{Šimon Trlifaj\footnote{Corresponding author at \url{Trlifaj_Simon@phd.ceu.edu}, Central European University, Vienna, Austria. Thanks to Cristina Corduneanu-Huci, Michael Dorsch, Mihaly Fazekas, Yian Yin, Shane M. Greenstein, Heesang Ryu, Joachim Henkel, Minyuan Zhao, Paul Heidhues, Jo Seldeslachts, Pierre Fleckinger, Mimra Wanda, Reinhilde Veugelers, Jannis Kück, Dirk Czarnitzki, David Cerero-Guerra, Agnes Batory, Bence Hamrak and participants of the 2024 Competition and Innovation Summer School, 2025 CEU Annual Doctoral Conference, 2025 Munich Summer Institute, and 2025 International Conference on Science of Science and Innovation for valuable feedback. Special thanks to Julius Knapp, Dale N. Hatfield, Reza Biazaran, Krista Witanowski and John Soscia. This project has been supported by grants from the Czech Fulbright Commission and the Central European University. Its contents are the sole responsibility of the author and may not represent the views of these institutions. All data and code necessary to replicate the results are available at: \url{https://doi.org/10.5281/zenodo.15131259}}
}
\date{\today{}}
    
\begin{document}

\setstretch{1.5} 


\maketitle


\begin{abstract}

    \noindent We exploit a 1998 administrative quasi-experiment to analyze the effects of government transparency on follow-on innovation. Using the universe of U.S. wireless products, we find that the publication of technical details of new products online increased follow-on technology use from an average of 0.2 to 1.36 subsequent products within five years. The increase peaked at three years, concentrated among competitors and incumbents, and was largest among foreign firms that used technologies introduced by U.S. firms. The results document that government regulations affect innovation by altering information frictions.

\end{abstract}


\section{Introduction}

Information frictions are central to theories of innovation. In Schumpeterian models of growth, the ease of imitation, driven in part by information frictions, determines the pace of innovation and the dynamics of market competition \citep{aghion_model_1992, aghion_competition_2001, acemoglu_innovation_2018}. In public economics theory, information frictions determine the inventor's ability to recover the costs of invention and govern the process through which innovation enters public knowledge \citep{nelson_simple_1959, arrow_economic_1962, anton_little_2004}. Balancing innovation disclosure with protection of the inventor is one of the central questions of innovation policy \citep{hall_choice_2014, bryan_innovation_2021}.

Empirically, the magnitude of the effects of information frictions at invention level is less well understood. Most of the literature on invention-level disclosure focuses on the patent system \citep[e.g.,][]{cox_patent_2019, gross_hidden_2023}, in part because undisclosed inventions are, by design, difficult to observe. Recent contributions focus on academic science, public infrastructure, or individual inventor mobility \citep[e.g.,][]{bryan_impact_2021, berkes_knowledge_2024, johnson_innovation_2024}, but the impact of information frictions on product innovation remains difficult to quantify.

This gap is notable given the significance private firms give to secrecy. Surveys consistently find that firms appropriate their inventions using secrecy and lead time significantly more than patenting across almost all industries. Firms that do patent do so endogenously \citep{kankanhalli_paradox_2024, hall_choice_2014}, skewing toward specific industries and large firms \citep{levin_appropriating_1987, arundel_relative_2001, cohen_protecting_2000}. In a 1994 U.S. survey, \citet[][p. 33]{cohen_protecting_2000} find that R\&D laboratories in the communication equipment industry report patents as an effective appropriation mechanism for 26 percent of their products (midpoint mean, N 34), compared to 47 and 66 percent for secrecy and lead time.

This paper investigates the effects of an exogenous reduction in information frictions on follow-on innovation by exploiting an administrative quasi-experiment in the U.S. wireless technology market. In 1998, the Federal Communications Commission (FCC) launched a public online database containing detailed technical exhibits of wireless-capable products on the U.S. market. The transition to this digital system created a bureaucratic friction during the uptake period: the FCC bulk-scanned exhibits of some products but not of others. Building on historical records and personal communication with FCC leadership and an implementing contractor, we analyze this friction in a quasi-experimental framework, where we compare the follow-on use of technologies introduced in products affected by this transparency shock against a control group of products that remained undigitized.

Using novel data on the universe of all wireless-capable products on the U.S. market \citep{greenstein_wireless_2026, kim_open_2024}, we estimate that the publication of exhibits increased follow-on use of technologies almost seven times, from a baseline of 0.2 average follow-on products to 1.36. This estimate incorporates time and category controls, in addition to maintaining balance across a variety of observed covariates that capture appropriation mechanisms, firm location and entry, and technology. It also passes two placebo tests to rule out unobserved product-level and firm-level heterogeneity.

To explore the mechanisms that drive the increase, we decompose the follow-on use by geography, product type and firm types. We find that follow-on innovation peaked three years after the focal product's application date. It most significantly increased follow-on use among non-U.S. firms of technologies in products introduced by U.S. firms, and it also increased use among U.S. firms, suggesting cross-country innovation spillovers (the analysis of technologies introduced in non-U.S. products is inconclusive). It concentrated among products in the same class as the focal product, but also increased follow-on use out of class, suggesting the use of treated technologies in new product types. Finally, it significantly increased follow-on use among competitors and incumbent companies (the effect on follow-on use by originator is inconclusive).

We next examine the response of the originator firm. To our surprise, we find no evidence of increased use of innovation appropriation mechanisms \citep{hall_choice_2014}, or of the originator engaging in escape competition through introducing further products with new technologies or different product classes \citep{aghion_competition_2001}. We do find weak evidence of an increased use of the same technology in the next product, signaling a pivot toward incremental innovation. These results, however, are limited by many firms being exposed to the transparency shock when introducing another product after the focal period.

This paper develops literatures on secrecy, disclosure and innovation, knowledge spillovers and market competition, and government regulation and innovation. Various empirical studies analyzed the effects of policies that either constrained \citep{gross_hidden_2023, rassenfosse_patents_2024} or expanded \citep{hegde_patent_2018, kim_innovation_2021, baruffaldi_patents_2020, hegde_patent_2023, cox_patent_2019} disclosure in patents on innovation. Other work leveraged plausibly exogenous administrative variation to identify the effects of market exclusivity \citep{gilchrist_patents_2016,sampat_how_2019}, the effects of open-access mandates in academic science \citep{bryan_impact_2021, staudt_mandating_2020, probst_impact_2023}, and the effects of non-compete agreements \citep{johnson_innovation_2024,samila_noncompete_2011}. Our contribution is in analyzing a permanent reduction in information friction in affected products, with minimal lag from market entry of the disclosed invention, on a population of products irrespective of their patenting status. In doing so, we contribute to the literature that develops alternative measures of innovation \citep{moser_how_2005, argente_patents_2021, sampat_how_2019}.

Prior literature on innovation spillovers analyzes geographical variation in access to information \citep{ganguli_paper_2020, furman_disclosure_2021, berkes_knowledge_2024}, including access to the internet \citep{forman_wires_2012, ertugrul_knowledge_2024}, and in-person interactions \citep{atkin_returns_2022,andrews_bar_2019,jin_face--face_2024}. Here, we analyze a shock where technical knowledge became accessible globally via the internet, leveraging between-product variation in exposure. We also contribute to works on the second-order effects of regulatory changes on innovation \citep{budish_firms_2015, michelman_sex_2024, klapper_entry_2006}, and expand the study of government transparency policies beyond the traditional focus on public-sector accountability \citep{porumbescu_government_2022, corduneanu-huci_freedom_2024}.

Governments have long shaped the information environment in which firms innovate, by restricting labor mobility in medieval Venice \citep{amato_island_1997}, paying for physical exhibits of inventions in the 18th-century Britain \citep{burrell_parliamentary_2015}, or publishing patent applications earlier since 1999 in the U.S. \citep{hegde_patent_2023}. Today, the pre-approval of artificial intelligence models, algorithmic transparency of social networks, or closures of U.S. government websites \citep{gotfredsen_fighting_2025} are policies that may affect innovation through altering information frictions. Here, we present novel evidence that these effects, even if unintended, may be large.

The paper is organized as follows. Section \ref{sec:background} details the institutional background of the FCC wireless product authorization program and describes the data. Section \ref{sec:empirical} outlines the empirical strategy and addresses the assumptions of the quasi-experimental setup. Section \ref{sec:results} presents the main results, and Section \ref{sec:extensions} explores the mechanisms driving follow-on innovation. Section \ref{sec:conclusion} concludes.

\section{Institutional background and data}
\label{sec:background}

We study follow-on innovation using a unique setting in the U.S. wireless industry. The 20th century saw a dramatic increase in the introduction of various products operating or intervening in the electromagnetic spectrum. To ensure safe and reliable operation of these devices, Section 302 of the Communications Act tasked the Federal Communications Commission (FCC) with regulating the industry \citep{knapp_considerations_1997,jackson_unlicensed_2009}. A major regulatory instrument in this effort is the product authorization program, which requires any firm marketing a product that uses or could interfere with wireless signals to comply with strict technical rules and document compliance in an application to the FCC, which then grants authorization for product marketing.

The wireless technologies which these products use can be broadly categorized into \emph{licensed} (where the operator needed a license from the FCC, such as cell towers, pagers, and TVs), \emph{unlicensed} (where no such license was needed, such as baby monitors and garage door openers), and \emph{unintentional} (where wireless signals are a byproduct, such as some computer peripherals and power tools), with various detailed regulatory requirements for each \citep{jackson_unlicensed_2009, milgrom_case_2011,greenstein_wireless_2026}. Today, many electronic products of daily use fall under this regulation, including smartphones and personal computers, wireless headphones, home appliances, and Wi-Fi routers, representing the frontier of high-tech consumer goods.

\subsection{The wireless products database}

One of the outcomes of the product authorization process is a public database of all wireless products on the U.S. market, available at \url{https://www.fcc.gov/oet/ea/fccid}. The database contains basic information such as applicant details, product classification, product description, and a list of operating frequencies. It also contains \emph{exhibits}: technical attachments in PDF format that document that the product meets the regulatory standards, including external and internal photos, block diagrams, schematics, parts lists, and test reports. These documents became a valuable source of information about new wireless products for journalists and enthusiasts following the development of wireless devices. In one recent example, the technology website The Verge reported on new Apple products having dormant wireless capability the company had not yet announced, based on information from the database \citep{tuohy_theres_2026}. Similar cases appear periodically in the technology press \citep[][e.g.]{weatherbed_whats_2024, gadgets360_xiaomi_2015}.

There is a wealth of blog posts online using the FCC database to learn about frequencies, wiring schematics, and components used in products listed in the database for hacking and reverse engineering wireless products. In appendix \ref{apx:sec:blogs} we summarize seven such cases. In one example, \citet{blake_reverse_2015} reverse-engineers a wireless BBQ thermometer to have it send temperature readings to a phone, writing `I also searched FCC filings using the [product's] FCC ID of N9ZMAV221 [...] which turned out to be extremely useful. In addition to verifying what I saw on the spectrum analyzer, I was able to download schematics and all sorts of other helpful information about the transmitter.'  \citet{laplante_reverse_2019} uses information from test reports from the FCC database in a process of reverse-engineering a mattress remote. They then sell a device that connects the remote to a smart home in an online shop.\footnote{Archived on May 21, 2025: \url{https://web.archive.org/web/20240420074714/https://www.tindie.com/products/cplaplante/temperbridge/}} These publicly documented reverse engineering projects often use the database in early stages to learn the key parameters of the device before dismantling it physically. As \citet{augusto_robbing_2021} notes: `As a new analyst [...], I was often told to check the FCC database before performing any hardware analysis, and for good reason.' A commercial website, \url{https://fccid.io} runs a copy of the database with advanced functionality against advertisements.

Some exhibits can be claimed as confidential by the firm, including schematics, block diagrams, operational descriptions, parts lists, and tune-up information (and, in rare cases, internal photos and the user manual). Even if this is so, many of the projects listed above still find useful information from exhibits that are public, such as test results and internal and external photos. Together, these examples suggest that the database contains information that may affect the adoption of specific wireless technologies not only by enthusiasts, but also by private firms. In order to analyze if this is the case, we next look closely at the time period at which the database was launched.

\subsection{The 1998 transparency shock}

The database of wireless products had not always been available online. In 1998, the FCC implemented two major changes to streamline the authorization process. These changes were presented at a 1997 IEEE conference \citep{knapp_considerations_1997}, shown for public consultation \citep{federal_communications_commission_fcc_1997, federal_communications_commission_fcc_1997-1} and enacted in two orders \citep{federal_communications_commission_fcc_1998, federal_communications_commission_fcc_1998-1}. The following recount of these changes is based on these documents, together with email correspondence with current and former FCC staff and a contractor who worked as a senior web architect on the project.\footnote{Personal email correspondence, March 2024 to April 2026.}

The first change made by the FCC in 1998 was to significantly lower the regulatory requirements for some products, mainly \emph{unintentional} radiators that interfered with wireless signals unintentionally. For these devices, the FCC would stop requiring an authorization prior to market introduction, and would only require a declaration of conformity based on an assessment of an accredited laboratory \citep{federal_communications_commission_fcc_1998-1}. We do not focus on these devices in the present analysis, other than ensuring they are excluded from the analysis to avoid contamination (appendix \ref{apx:fig:deregulation}).

In a second change, the FCC digitized the authorization application process and launched the above-referenced website with the database of FCC products. The stated purpose of the decision was for the process to be `streamlined', enabling parallel processing of applications and improving the accuracy and speed of the application review process. The justification of the changes also included supporting the introduction of innovative products to the market, but only in the context of the deregulation, not the digitalization \citep{knapp_considerations_1997,federal_communications_commission_fcc_1998-1}.

Although the digitalization was largely well received by the industry, some firms raised concerns about the confidentiality of the digital filings \citep{federal_communications_commission_fcc_1998-1}. As the FCC made clear in a public consultation process, the policy shock did not change the legal regime of the authorizations or the information requirements \citep{federal_communications_commission_fcc_1998}. The same scope of information that became available online had been under a public access regime before the digitization. However, accessing the information prior to the change meant going to the FCC's offices or filing individual requests under the Freedom of Information Act. Although the FCC provided such information routinely, it meant significantly higher costs and effort, especially for those seeking the information from abroad. 

Under the new application procedure, all applications had to be submitted digitally. The digital submission system launched in March 1998, and the FCC allowed for a one-year uptake period during which applicants could choose to file either on paper or digitally \citep{federal_communications_commission_fcc_1998-1}. The FCC also hired a contractor to implement the system and digitize basic information about all applications going back to 1981. As a result, basic information (applicant details, product classification, product description, and a list of operating frequencies) became available for all products, whereas exhibits (such as external and internal photos, block diagrams, schematics, parts lists, and test reports) were available only for products in the new digitized system. For applications filed in the new system, the information on the most recent wireless-capable products became instantly available for anyone with internet connectivity, from the moment of market entry.

The uptake was fast: within just eight months, virtually all applications were submitted digitally. One reason for this was that the FCC appears to have scanned and digitized some of the applications during the uptake period on top of those voluntarily submitted digitally by the applicant, a possibility we will explore after introducing the dataset.

\subsection{Data construction}

We construct the dataset by web-scraping the FCC product authorization database, which contains the universe of all wireless devices introduced to the U.S. market since 1981. We start with firm identifiers (grantee codes), which the FCC issues to companies and publishes on its website.\footnote{Available at \url{https://apps.fcc.gov/oetcf/eas/reports/GranteeSearch.cfm}} As one firm may have registered multiple identifiers, we clean the name of each company and merge all firms with the same cleaned name and country into one entity. The FCC ID search has a limited API interface which allows us to download all products registered by each firm based on its ID.\footnote{\url{https://web.archive.org/web/20260421123250/https://apps.fcc.gov/kdb/GetAttachment.html?id=a2TAM\%2Fce88o0NtvDeTzB\%2Bg\%3D\%3D&desc=953436\%20OET\%20Laboratory\%20Services\%20API\%20v04&tracking_number=50070}} By querying the API with all grantee IDs, we end up with a list of all products in the database, identified through their unique FCC IDs. We encode \emph{NewEntrant} equal to 1 for the first product of an applicant (13 percent of products with new frequency in five years before and after the 1998 policy change, excluding deregulated products, see Table \ref{apx:tab:summary} for full summary statistics).

For each product, we web-scrape additional information from the web interface of the FCC database.\footnote{\url{https://www.fcc.gov/oet/ea/fccid}} Some products can have more than one equipment authorization application associated with them, as the technologies contained in them may be subject to multiple different rules, and the FCC allows for applications that change the original product. Because these multiple applications still describe one product, we aggregate information on the \emph{FCCID} level based on the original application. We encode \emph{Changed} equal to 1 if the product was changed at some point from the first application (10 percent of products).

Two forms associated with each application that are available for all products: the application form (Form 731) and the equipment authorization grant form. We prefer information from the grant form when available, as it has been validated by FCC staff. From here, we encode the \emph{ApplicationDate}, the \emph{GrantDate}, the rule parts and the product classes. Rule parts list the specific technical regulations the product needs to meet, and from them we identify productss deregulated under the 1998 regulatory change which we exclude from the dataset (details in appendix \ref{apx:sec:deregulated}). Classes list the technological product categories gradually added by the FCC as new product types were introduced. Although informative, these classes are highly imbalanced, with some containing just a few products in the period of interest, and also because some products may fall under more than one class. We aggregate classes in two ways: we encode \emph{LicensedClass} equal to 1 for products with at least one class indicating they use licensed technology (53 percent of products). We also construct and classify each product into seven broader \emph{Categories} based on classes with at least five percent of products (see appendix \ref{apx:sec:category}). 

We encode \emph{Secrecy} equal to 1 if the applicant claims confidentiality on the exhibits in the application form.\footnote{Form 731 distinguishes between short- and long-term confidentiality. We disregard short-term confidentiality because it is highly correlated with long-term confidentiality and is limited to 180 days and at most the market entry date. The applicant must justify the request in a letter and pay a fee that, in 1998, amounted to 130 dollars \citep{federal_communications_commission_fcc_1998}. Applications classified for national security reasons are under a stricter regime, and are not available in this dataset here.} 54 percent of products claim secrecy, with a significant increase after the policy shock: five years after the change, in 2003, 71 percent of applicants claimed secrecy (see Figure \ref{apx:fig:secrecy_patenting}). As a coarse measure of intellectual property protections, we use a semantic matching algorithm to match firms to USPTO / PatentsView data by cleaned name. We only consider U.S. patents even for foreign applicants, because the U.S. is the place of marketing of products under FCC regulation, and any potential infringement falls under the jurisdiction of the USPTO. \emph{RecentPatent} and \emph{FuturePatent} are binary variables equal to 1 if the firm applied for at least one patent within five years prior (post) to the FCC application (62 and  64 percent, respectively).

\subsubsection{Frequencies as technologies}

To study follow-on innovation, we need to identify products that introduced new technologies and track the use of these technologies across multiple products. To that end, we use information about the frequencies used in each product, listed in the grant form by upper and lower bound and output. This approach builds on earlier work which used technological properties of patents and products to measure how they relate to one another, such as specific gene sequences in medical technologies \citep{sampat_how_2019}, chemical structures in drugs \citep{wagner_mapping_2022}, pests and pathogens in agriculture \citep{moscona_inappropriate_2025}, product characteristics in retail products \citep{argente_product_2022} or patent applications in a narrow patent class \citep{bell_who_2018, rigby_technological_2015, hall_nber_2001}. Similar to these approaches, we use product characteristics to track the use of specific technology combinations from the first product throughout following products.

We begin by representing each product by the frequencies it uses, defined by lower and upper bounds (in megahertz) and outputs (in watts). We then encode \emph{\textit{NewFrequency}} as a binary variable equal to 1 if a product uses a combination of frequencies not used by any prior product (30 percent of products), a proxy for the use of a new technology. Products with a new frequency combination are 1.3 times more likely to have a new product description, and five times more likely to be in a new product class around the focal period. 

We next track other products using the same frequency combination as the first focal product. The assumption underlying this approach is that a product building on the same technology would be more likely to use the same frequency as the original product. For wireless products, the operating frequency constitutes its fundamental engineering feature, one that can be replicated with access to key documents such as schematics or block diagrams. The reverse-engineering examples described in appendix \ref{apx:sec:blogs} often start by analyzing the frequencies used by the target product, and if the purpose of the reverse engineering is to build competing or complementary products, it is more likely they would use the same frequency.

Appendix \ref{apx:sec:examples} shows examples of two products in the FCC database, the frequencies they used, and the characteristics of other products that used the same frequency combinations. One of them is a crane remote controller (FCC IDCBFCRANET1) applied for on October 5, 1998. Seven other follow-on products use the same frequency in the next five years, including an industrial remote controller introduced by a different company three years later (ONF-8516RX840), with similar design.

As the primary outcome variable, we count the future products that used the same combination of frequencies within 5 years after the focal product's application date, \emph{ForwardUse} (mean 1.06, standard deviation 5.34). To avoid capturing concurrent multi-product launches by the same firm or co-inventions, we exclude products applied for within the first year. Inversely, we define \emph{BackwardUse} as the count of products that used the same frequency until the focal product's introduction (mean 134, standard deviation 288). As is typical for count data in innovation literature, both variables are heavily right-skewed.

We count separately the future frequency use by the originator firm that applied for the first product (\emph{ForwardUse, Originator}, mean 0.08, standard deviation 0.56) and by other firms (\emph{ForwardUse, Competitor}, mean 0.98, standard deviation 5.21). We also split the count between products by incumbents and new entrants (\emph{Incumbent}, \emph{Entrant}), products in the same class as the original product and in a different class (\emph{InClass}, \emph{OutClass}). Depending on whether the follow-on product was introduced by a country that joined the World Trade Organization (WTO) prior to 1998, we further split the count by domestic, foreign WTO and foreign non-WTO firms (\emph{Domestic}, \emph{Foreign (Non-)WTO}), and the combination of these splits.\footnote{Dates of WTO accession taken from \url{http://web.archive.org/web/20251021112700/https://www.wto.org/english/res_e/booksp_e/sli_e/4wtomembers.pdf}.}

We also use frequencies to identify products with internet connectivity. We encode \emph{Internet} as a binary variable equal to 1 if the product uses a frequency that is associated with 3G, 4G or Wi-Fi internet connectivity (23 percent, details in Supplemental appendix \ref{apx:sec:internet}). We also encode \emph{FrequencySimple} for products with a single frequency combination. Together, the dataset gives us a detailed picture of wireless products and the nature of follow-on use of the wireless technologies contained in those products around the policy shift.

\subsection{Who scanned the exhibits?}

\begin{figure}[t]
    \centering
\includegraphics[width=0.9\textwidth,keepaspectratio]{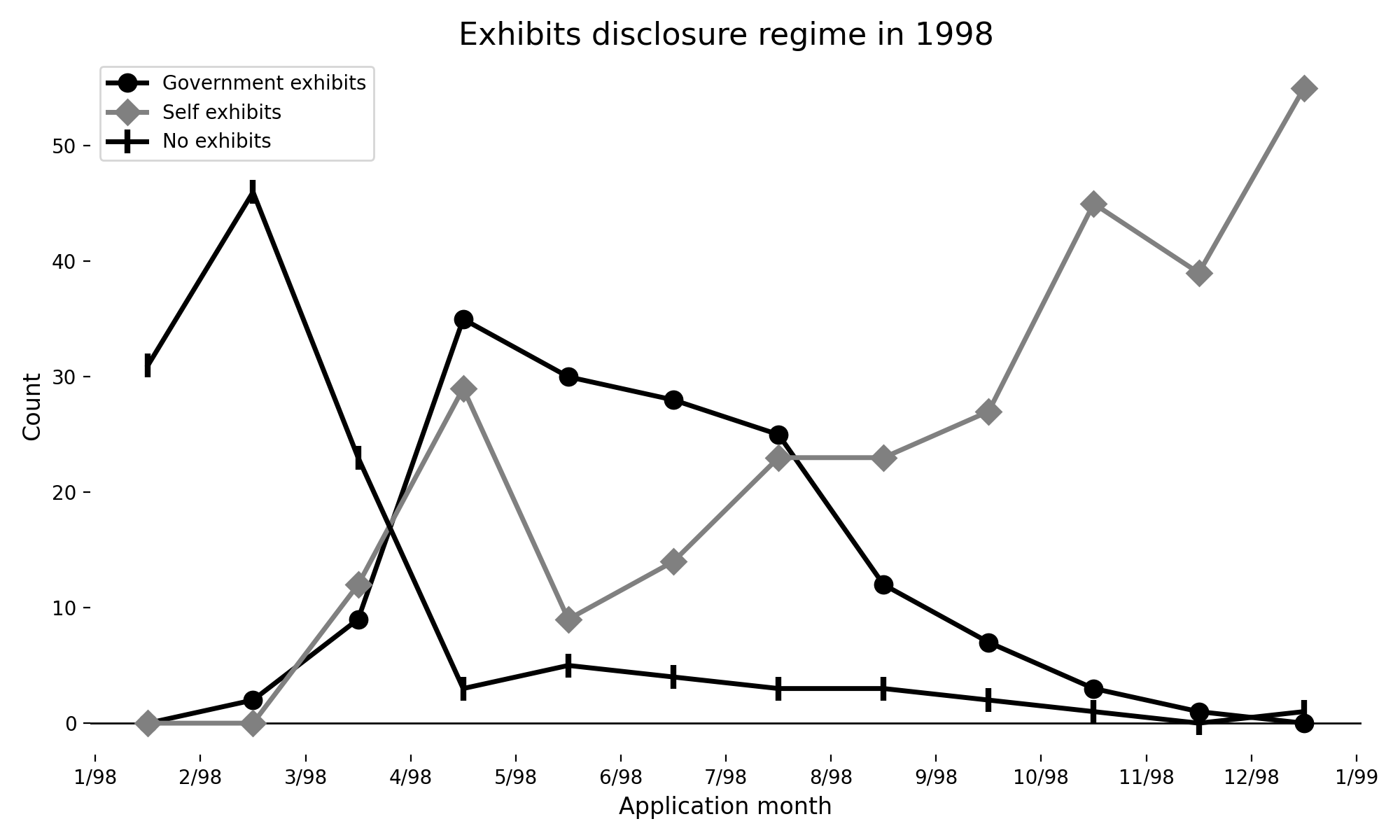}
    \caption{Proportion of products under different exhibits disclosure regimes.}
    \label{fig:exhibits_availability}
\end{figure}

Our empirical strategy rests on the classification of products in 1998 into three different exhibits disclosure regimes: those submitted digitally by the applicant (\emph{Self exhibits}), those submitted on paper and scanned by the FCC or its contractor (\emph{Government exhibits}), and those submitted on paper and not scanned (\emph{No exhibits}). Distinguishing products with no exhibits is straightforward: their exhibits list is simply unavailable in the database. Distinguishing products scanned by the government and by the applicant during the application process requires a more careful analysis.

The FCC proposed it might scan applications in its 1998 decision, stating `[we] encourage [applicants] to scan their attachments as PDF files and submit them electronically. If [they] cannot do so, we will attempt to scan the non-electronic portion of the filing [...]' \citep[p. 11,333][]{federal_communications_commission_fcc_1998}. We verified this was indeed done with the senior web architect of the digitization, who confirmed that contractors were hired to 'bulk-scan legacy paper applications' during the transition period.

As we could not find a direct indicator of whether the FCC or their contractor scanned the application in the database, we construct a proxy leveraging the timestamps on the exhibits list to identify productss that were likely scanned by the FCC or their contractor. To that end, we compare the date the exhibits were submitted to the electronic system (`submission date') and the authorization grant date. 

For virtually all products authorized after the full electronic system rollout, the timeline follows a chronological order: the exhibits submission date precedes the authorization grant date. We interpret this as the applicant uploading the exhibits to the system, either as part of the initial submission or during the authorization process, and the FCC then granting the authorization and making the exhibits publicly available. However, for a significant proportion of products in 1998, the order is reversed: the grant date precedes the exhibits submission date. This occurs frequently in 1998 and almost never later. For these applications, we believe the later submission date captures the scanning of the exhibits into the database by the FCC or its contractor. We classify products for which the grant date precedes the submission date into the government exhibits disclosure regime.\footnote{This is a conservative definition. Some products have the exhibits submission date before the grant date but after the application date.  We confirmed with the senior web architect that during the transition, contractors occasionally scanned pending paper applications concurrently during the review process, resulting in the submission date preceding the grant date. We cannot distinguish if this scanning was initiated by the applicant of by the FCC. It may reflect additional exhibits being submitted to the application at the request of the FCC or at the request of the applicant. Requiring that the digitization occurred after the product grant excludes these possibly endogenous cases.}

Figure \ref{fig:exhibits_availability} plots the count of the three disclosure regimes of products during 1998. It shows that at the beginning of the year, virtually all products had no exhibits scanned, and by the end of the year, virtually all products had exhibits scanned by the applicant. For products applied for after the system launch in March, the government retroactively scanned a significant proportion of exhibits, but their proportion decreased with the increasing proportion of self exhibits, dropping to virtually none in the following year. Between March and October, 442, 429, and 153 products in the self, government and no exhibits disclosure regimes, respectively, were submitted for authorization.

To validate this measurement approach, we check the weekday distribution of the exhibits submission date to test if those exhibits that were presumably scanned by the FCC or their contractor were less likely to be scanned on the weekend. As Figure \ref{apx:fig:exhibits_weekday} shows, about 3.5 percent of self exhibited products in 1998 were submitted on the weekend, compared to \emph{no} products in the government exhibits group. The distribution within the work week is also more uniform for the government exhibits disclosure regime. We believe this pattern corresponds to the scanning occurring during opening hours of the FCC premises and a steady government administrative workload, in contrast to the self exhibits being submitted on the weekend and unevenly distributed in the week.

We further validate this timeline proxy by examining the embedded metadata of the digitized exhibits. In \ref{apx:sec:exhibits}, we examine a random sample of product applications from both the self exhibits and government exhibits group, ten of each. The government-scanned exhibits show uniform encoding software (Acrobat Distiller 4.0 for Windows) and consistent authors, whereas self-submitted exhibits display high variance in both metadata values. ANother alternative source of the mismatch in dates is that the FCC scanned these documents when the applicant filed a secondary application with changes to the original product. As we document in the next section, there is no difference in the proportion of changed products between the no exhibits and government exhibits groups. Examining the correspondence attached to these applications, we also found no evidence of the exhibits being submitted after the grant date.

The senior web architect confirmed our approach, stating that the patterns of the grant date preceding the exhibits submission date and of metadata information are `consistent with post-hoc digitization of legacy records during system migration'.

Using this distinction between the three disclosure regimes, we next use the government exhibits group as our quasi-experimental treatment and the no exhibits as a control groups. In the next section, we examine if the selection of products that were scanned by the government was exogenous to the strategic choice of the inventor to self-disclose and to follow-on innovation.

\section{Empirical strategy}
\label{sec:empirical}

We use the institutional setup of the FCC transparency shock to analyze the effects of government transparency on follow-on use of new technologies. We compare products in the no exhibits group (control) and the government exhibits group (treatment), treating the assignment into scanning as a quasi-experiment with controls for time and category. Our main sample focuses on the period during which both treatments occurred, Match to October 1998, and excludes deregulated products, products with previously used frequency combinations, and self-exhibited products.

The main model estimates the effect of the exhibits being scanned (\emph{GovernmentExhibits}) on the follow-on use of a wireless technology:

\begin{equation} \label{eq:main_spec}
    \eta_i = \beta \text{GovernmentExhibits}_{i} + \rho \text{ApplicationDate}_{i} + \mathbf{X}_{i}^{\prime}\theta + \delta_{c(i)}
\end{equation}

The dependent variable, $\text{ForwardUse}_i$, is the count of subsequent products that use the same frequency combination as focal product $i$ within the relevant follow-on window. Because the outcome is non-negative and right-skewed, we estimate the model using Poisson pseudo-maximum likelihood (PPML), so that the conditional mean satisfies $\mathbb{E}[\text{ForwardUse}_i \mid \cdot] = \exp(\eta_i)$ \citep{silva_log_2006,chen_logs_2024}. The treatment variable, $\text{GovernmentExhibits}_{i}$, indicates whether the product's exhibits were scanned and published on the government website. $\text{ApplicationDate}_{i}$ is a continuous time control, $\mathbf{X}_{i}$ includes product- and firm-level controls for new entrants, U.S. applicants, secrecy, recent patenting, and internet technologies, and $\delta_{c(i)}$ denotes product-category fixed effects.

We are primarily interested in $\beta$, which estimates the effect of exhibits availability on the follow-on use of wireless technologies contained in the exposed product. We further report log-linear OLS and binary models as robustness checks for alternative margins of the outcome. We clustered standard errors by applicant, as multiple products may be submitted by the same firm and treatment assignment may be correlated within applicant.

We also develop a set of alternative outcomes to analyze the effect of the transparency shock on the original firm. \emph{Survival} is a binary variable indicating if a firm introduced at least one product five years or later after the focal product. \emph{NProducts}, \emph{NProductsNew}, \emph{NProductsSecrecy}, \emph{NProductsPatent} counts the number of products the originator firm introduces two to five years after the focal product; all products, products with a new frequency combination, protected with secrecy, and with a recent patent, respectively.\footnote{We exclude the first two years to allow for a lag between treatment and change in behavior, and to avoid counting products within the treatment period in the extended sample, as explained later.}

\subsection{Causal identification}

For $\beta$ to be interpreted causally, products selected into government scanning must not differ systematically from control products in other factors that affect follow-on use, conditional on the observed controls \citep{imbens_causal_2015}. As this assumption is not testable directly, we present the institutional record, observable balance, and robustness exercises to assess its plausibility indirectly \citep{imbens_causal_2015, austin_balance_2009}.

We exclude self-exhibited products because voluntary digital submission is plausibly endogenous to product and firm characteristics associated with follow-on innovation. The voluntary decision to submit scanned documentation through the new online system might be confounded by other factors that affect the outcome. For example, a more innovative firm might be more likely to use the new digitized submission system, and may also be more likely to introduce products with technologies more attractive to follow-on use.

Our focus is on products in the no exhibits and government exhibits group. Between these two groups, identification requires that, conditional on the included controls, scanning was unrelated to other determinants of follow-on use. This condition might be violated, for example, if the FCC or their contractor chose to prioritize applications from new or domestic firms. Such a pattern would introduce a bias in our estimates of the effect of the scanning on follow-on use.

Neither FCC staff at the time nor the the contractor recall that a set of product authorization applications would be systematically prioritized in the government scanned group. As the contractor recalls, the `[new system] was generally executed in a phased manner, with earlier submissions digitized in bulk and then loaded into EAS as part of the initial system population. The workflow was largely ``first in, first out'' for backlog processing. During the transition period, sequencing varied across application types and contractor support cycles.' This institutional evidence is consistent with scanning not having targeted technologically promising products, but it is insufficient on its own. We next examine balance in observable characteristics, especially time and product category.


In time, the two groups are clearly imbalanced. As Figure \ref{fig:exhibits_availability} shows, most products were not scanned in March 1998, but this flipped between April and October, when more products were government scanned than were not. One potential concern is that follow-on use was evolving over the same period for reasons unrelated to exhibits availability. In the main specification, we therefore include a linear control for application date. In [robustness], we further allow for more flexible time adjustment using application-month fixed effects and a quadratic time trend. One possibility is that March, the month of the announcement and also the most diverging treatment balance, is different from the rest of the period in ways that confound the outcome. In [robustness] we present results excluding March observations. Time could also confound the results due to seasonality. In [robustness], we include results from a seasonal placebo test, in which treatment is randomly reassigned in pre-1998 data to match the monthly treatment distribution observed in 1998 \citep{young_channeling_2019}.

\begin{table}

    \caption{Covariate balance.}
    \centering
    \begin{tabular}[t]{lcccccc}
    \toprule
     & No exhibits & Government exhibits & $\delta$ & SE & P\\
    \emph{Variable} & (1) & (2) & (3) & (4) & (5) \\
    \midrule
    RecentPatent & 0.52 & 0.48 & -0.07 & 0.11 & 0.53\\
    Secrecy & 0.30 & 0.27 & -0.04 & 0.09 & 0.63\\
    U.S. applicant & 0.68 & 0.61 & -0.09 & 0.10 & 0.36\\
    NewEntrant & 0.09 & 0.11 & 0.04 & 0.05 & 0.44\\
    FrequencySimple & 0.45 & 0.60 & 0.15 & 0.10 & 0.15\\
    Internet & 0.16 & 0.17 & -0.01 & 0.07 & 0.88\\
    Licensed & 0.61 & 0.36 & -0.28 & 0.10 & 0.00\\
    \addlinespace
    Changed & 0.09 & 0.10 & 0.00 & 0.05 & 0.95\\
    \addlinespace
    \addlinespace
    Number of products & 44 & 149 &  &  & \\
    \bottomrule
    \end{tabular}

    \footnotesize
    
    Covariates balance test. (1) mean value of not scanned products, (2) mean value of products scanned by the FCC or its contractor, (3--5) the time-trend adjusted coefficient, standard error, and p-value $\delta$ from raw $X_i = \alpha + \delta \text{GovtExhibits}_i + \rho \text{ApplicationDate}_i + \varepsilon_i$, where $X_i$ is the variable in the first column. Products with new frequency combinations between March and October 1998, excluding deregulated and self-exhibited products. Standard error clustered by Applicant. See \ref{apx:tab:balance} for Category-adjusted covariates balance.
    \label{tab:balance}

\end{table}

Table \ref{tab:balance} shows the covariates balance between the treatment and control groups \citep{austin_balance_2009}. It tests if the government was more likely to scan products with higher use of appropriation mechanisms ($Secrecy$ and $RecentPatent$), firm location and entry ($ApplicantUS$ and $NewApplicant$), technology ($FrequencySimple$ and $Internet$) and product licensed spectrum use ($CategoryLicensed$). We also include one post-treatment variable, $Changed$, to test if original applications of products that were later changed by the applicant were also more likely to be scanned.

The results show a significant imbalance in the proportion of licensed products. This is of concern, as products using a licensed spectrum face higher market-entry barriers \citep{jackson_unlicensed_2009, milgrom_case_2011}, and may therefore exhibit lower follow-on innovation due to these regulatory requirements rather than the exhibits treatment.

To examine if there is a broader imbalance between categories, Table \ref{apx:tab:treatment_category} estimates the probability of government exhibits by category types. It shows that the imbalance is concentrated in the two licensed products categories, while other product categories do not significantly differ from the baseline of unlicensed Part 15 products. The category imbalance would correspond to different sequencing by `application type' referenced by the contractor: licensed products appear to have been less likely to be scanned by the FCC or their contractor. In all our specifications, we control for product categories, and present results split by licensed and non-licensed products in appendix \ref{apx:robustness}. 

All other variables are well balanced, suggesting that the FCC or its contractor did not systematically prioritize applications along these observed dimensions. To rule out that this insignificant result is the result of a small sample or the known category imbalance, Table \ref{apx:tab:balance} presents balance with category fixed effects (panel A) and balance extended to a larger sample of products without a new frequency combination (panel B). Both extended samples show a consistent pattern: all observed variables are statistically similar between the treatment and control groups, except products with licensed frequencies. $FrequencySimple$, which is somewhat imbalanced in Table \ref{tab:balance} becomes well-balanced in these extended tests.

To compare the variance of these variables, we also include the self-exhibited products group in Table \ref{apx:tab:balance} (columns 3, and 7--9). Compared to the no exhibits group, these self-treated products are more likely to be applied for under secrecy, by firms that recently patented, by new entrants, and contain internet connectivity (panel A). On the extended sample, they are also likely to contain new frequency combinations and be changed later (panel B). This corresponds to products with more novel, sophisticated and protected technologies being self-scanned by the applicant---in contrast to those scanned by the government which show no such differences from the control group.

\begin{figure}[t]
    \centering
    \begin{subfigure}[t]{0.49\textwidth}
        \centering
        \includegraphics[width=\textwidth]{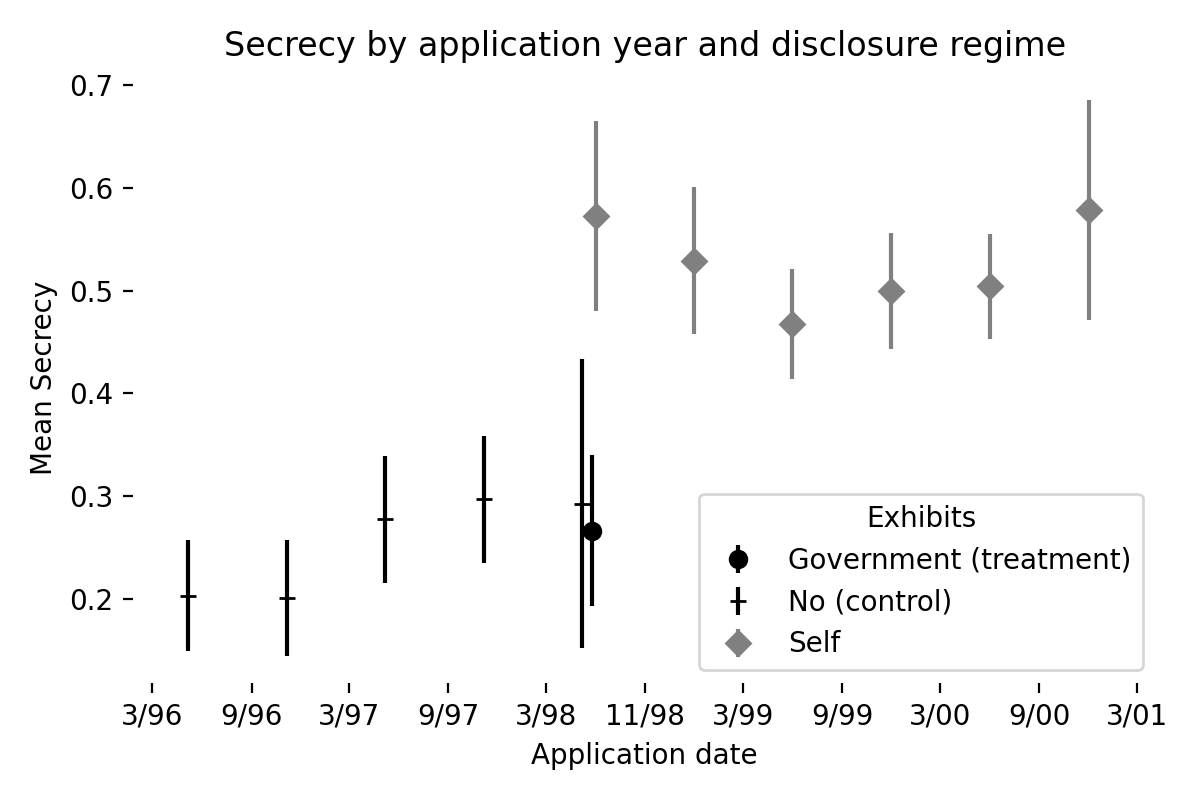}
        \caption{Secrecy by disclosure regime.}
    \end{subfigure}
    \hfill
    \begin{subfigure}[t]{0.49\textwidth}
        \centering
        \includegraphics[width=\textwidth]{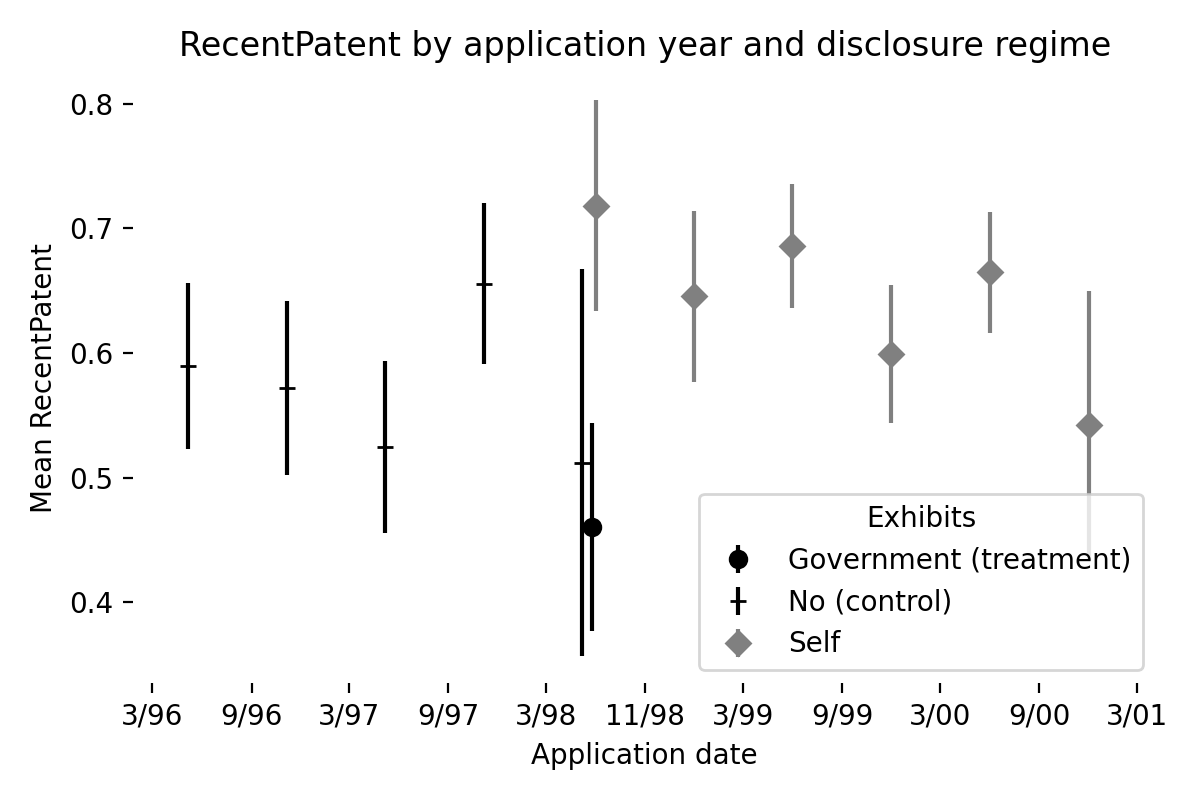}
        \caption{RecentPatent by disclosure regime.}
    \end{subfigure}
    \caption{Unconditional bi-yearly means, covariates.}
    \footnotesize Means of $Secrecy$ and $RecentPatent$ by disclosure regime in time, 95 percent confidence intervals. The time bins are by six months starting in March and September of each year. The March 1998 bin extends to October so it matches the treatment period. Groups of fewer than 20 observations are excluded for clarity. New frequency products non-deregulated products only.
    \label{fig:biyearly_controls}
\end{figure}

A potential threat to identification is the presence of contemporaneous macroeconomic or regulatory shocks, such as the late-1990s wave of trade liberalization (e.g., WTO expansions) or changes to intellectual property regimes such as the 1999 American Inventor's Protection Act \citep{hegde_patent_2023}. The quasi-experimental design mitigates this concern by relying on cross-sectional variation in follow-on use for products introduced in a narrow, eight-month transition window. Apart from the time imbalance discussed above and examined in robustness checks, both government-scanned and unscanned products were equally exposed to these broader structural shifts.

For a broader context in time, we show the evolution of these characteristics by disclosure regime. Figure \ref{fig:biyearly_controls} plots two innovation appropriatin prixy variables, $Secrecy$ (panel A) and $RecentPatent$ (panel B). Comparing means before and after the policy change, it shows a sharp increase in secrecy following the introduction of the digitized application system. This would correspond to the self exhibiting applicants being concerned about disclosure \citep{federal_communications_commission_fcc_1998-1}, or an increased ease of claiming confidentiality in the digital application system. We do not observe a similar increase in recent patents, which could be in part due to the construction of the variable, which considers patents by the firm in the prior three years, and in part due to the increase of new entrants, who patent less.

During the treatment period, the use of secrecy and patenting is notably higher among self-exhibited products, while the treatment and control groups maintain a lower and comparable proportion of each. This suggests that the applicants in the treatment group were unaware that their products would be scanned, as they did not seek legal protections in terms of secrecy and patenting. Appendix \ref{apx:sec:biyearly} presents analogous plots for other control variables.

These results show that the treatment and control are imbalanced in time and licensed product classes, and balanced along all other observed dimensions, even when we consider a broader sample of products. This is consistent with scanning a backlog of product authorizations in a process that differed by application date and category, but was similar in all other observed firm and product characteristics.

To further test treatment selection on unobserved characteristics, we perform two placebo tests on the firm and product levels. In the firm-level placebo, we subset the analysis to firms with at least one product introduced prior to the treatment period. We then replace the true treatment product outcome and controls with the last pre-treatment product, keeping the placebo treatment based on the firm's truly treated product. If $\beta'$ is significant in this specification, it would indicate the treatment prioritized products by firms with higher follow-on products irrespective of treatment, thus confounding $\beta$ in the main specification.

In a second placebo test, we expand the treatment and control groups to products with previously used frequencies. We then change the outcome variable from $\text{ForwardUse}_i$ to $\text{BackwardUse}_i$. If $\beta''$ is significant in this specification it would indicate the treatment prioritized highly used frequency combinations, again confounding $\beta$ in the main specification. As we report in appendix \ref{apx:robustness}, we find no significant result in either placebo test.

Taken together, these results are consistent with interpreting $\beta$ as the causal effect of exhibits availability, conditional on time controls and product class composition, even though, given the observational nature of our data, we cannot fully rule out unobserved confounding \citep{imbens_causal_2015}.
a
\subsection{Control group extension}

Our main sample, based on products from March to October 1998, yields a set of 193 products, of which 149 were government-scanned and 44 were not. This set is conceptually well-defined, but its size limits statistical power given the sparse distribution of the outcome. As the PPML estimate of $\beta$ relies on the control mean in the denominator $\exp(\beta) - 1 = \frac{\mathbb{E}[Y_1 \mid \mathbf{X}] - \mathbb{E}[Y_0 \mid \mathbf{X}]}{\mathbb{E}[Y_0 \mid \mathbf{X}]}$, a low number of non-zero observationss in the control group risks inflating the estimate \citep{cameron_microeconometrics_2005,king_logistic_2001}. For transparency on this issue, we report the number of non-zero observationss and unconditional mean outcome for treatment and control groups in fit statistics.

To expand the control group of not scanned products, we further include products since March 1997 in an extended set of 606 products, of which 151 were government-scanned and 455 were not.\footnote{Two additional products with application date in February 1998 were scanned by the government, which explains the addition of two products in the treatment group.} This extended set is necessarily imbalanced in time, but provides a larger set of observations for the control group.

This extension raises two concerns: it exacerbates the differences between the treatment and control groups in time, and it includes products in the control that would have been self-scanned by the applicant had there been the opportunity. We include linear time control in the main specification and a set of polynomial and month-fixed in robustness checks in appendix \ref{apx:robustness}. We also restrict the extension threshold to September 1997 and January 1998 in []. None of these significantly change the results. As Figure \ref{fig:biyearly_outcome} shows, the mean outcome from March 1997 to February 1998 is close to the mean of the March to October 1998 main control. It also shows the self-exhibited products mean above the no exhibits mean, suggesting the inclusion of would-be self-scanned products in this extended sample leads to conservative estimates: it would bias the control group mean upward and the coefficient $\beta$ downward. In Table \ref{apx:tab:balance} (panel C), we present coefficient balance on this extended sample, which retains good balance. This extended sample complements the conceptually cleaner but smaller control group.

\section{Does government transparency increase follow-on innovation?}
\label{sec:results}

\begin{figure}[t]
    
    \centering
    \includegraphics[width=.8\textwidth]{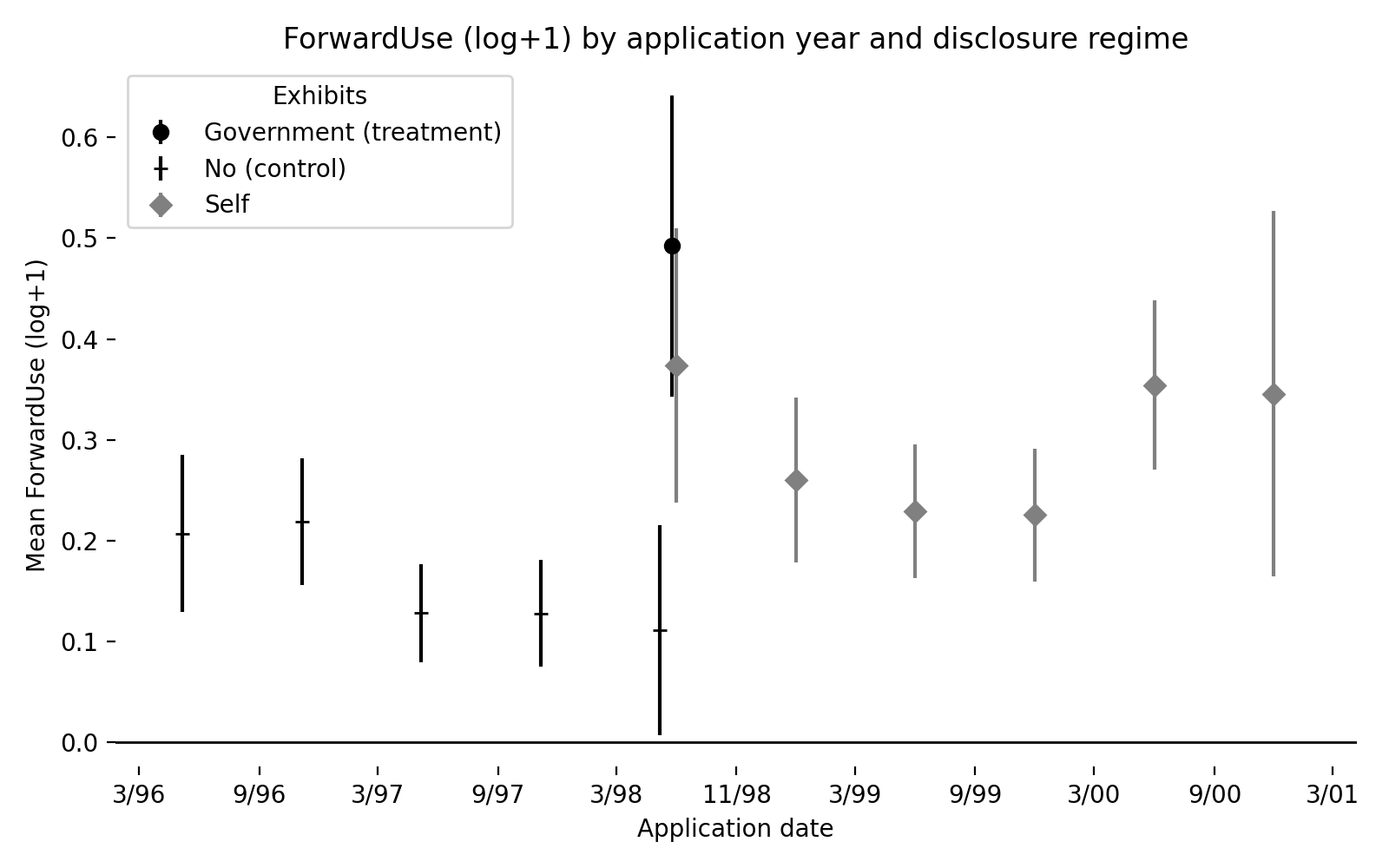}
    \caption{ForwardUse by disclosure regime, unconditional bi-yearly means.}

    \footnotesize Unconditional means of the main outcome variable, $ForwardUse (log + 1)$  by disclosure regime in time, 95 percent confidence intervals. The time bins are by six months starting in March and September of each year. The March 1998 bin extends to October so it matches the treatment period. Groups of fewer than 20 observations are excluded for clarity. New frequency products, non-deregulated products only.

    \label{fig:biyearly_outcome}

\end{figure}

We first present log-transformed unconditional means of the outcome in half-year intervals in Figure \ref{fig:biyearly_outcome}. New products introduced before the treatment period, which starts in March 1998, show significantly lower mean forward use than products introduced after October 1998. The most dramatic difference is during the treatment period, where frequency combinations introduced in products in the government exhibits group show eleven times higher use than those introduced in the no exhibits group. Self-exhibited products also see elevated levels, but this could be due to higher underlying value as signaled by increased use of secrecy and patenting in this group. In contrast, the government exhibits are comparable to the no exhibits group in most covariates, with the exception of time and licensed category imbalance. This pattern suggests that government transparency increased follow-on innovation, but the increase may be also due to other events happening at the same time.

\begin{table}
    \input{tables/result_main.tex}

    \caption{Effect of government transparency on follow-on innovation}
    \label{tab:main}

    \footnotesize PPML estimates of the effect of government-scanned exhibits on follow-on innovation (\emph{ForwardUse}). The main sample includes products with new frequency combinations submitted between March and October 1998, excluding deregulated and self-exhibited products. Extended control sample includes observations since March 1997. Columns (1) and (4) include baseline controls; columns (2) and (5) add \\emph{LicensedClass}; columns (3) and (6) include full category fixed effects. Standard errors in parentheses, clustered by applicant.

\end{table}

To isolate the effect and obtain a conditional causal extimate, we next fit the main PPML model in Table \ref{tab:main}. Column 1 shows the time-adjusted result, confirming a significantly higher forward use of frequency combinations introduced in the government exhibits group. To rule out this is due to the imbalance in licensed products discussed earlier, column 2 includes a binary fixed effect for licensed products. The coefficient retains significance, but drops in magnitude. This is consistent with non-licensed frequencies being exposed to higher forward use, and also being overrepresented in the control group.

Column 3 represents the main result with more granular category fixed effects. It estimates $\beta$ at $1.914$, which corresponds to frequencies introduced in products in the government exhibits group being used 6.78 times more often in other products within five years. This is slightly lower than the unconditional increase (column 1) and comparable to conditioning on licensed products (column 2). It represents an increase to 1.36 follow-on products compared to the unconditional mean of 0.2 products in the control group.

Results in columns 1--3 are based on the main sample from March to October 1998. One concern is that the control mean in this sample could be driven by just five non-zero observations (out of the 44 in the control group). Column 3 further drops three observations in the treatment group due to all zero outcomes in one of the product categories. We present results based on the extended sample in columns 4--6, which expands the non-zero control group observations to 58 (out of 455). We note that this does not change the control mean significantly, increasing it from $0.2$ to $0.26$. Conceptually, this may correspond to the inclusion of would-be self-exhibited products, which in the treatment period show higher average (plot \ref{fig:biyearly_outcome}).

The results on this extended sample are consistent with the main sample. The coefficient with linear time and category controls in column (6) corresponds to frequencies in the government exhibits group being 3.8 times more used in future products than those in the control group. 


We next examine the robustness of these results in terms of category, time and product- and firm-level unobserved characteristics. Table \ref{apx:tab:licensed} examines our main result split between products in licensed and non-licensed classes. It shows the results being mostly driven by an increase among non-licensed products on the extended sample.\footnote{We consider the extended sample more stable, as the main sample licensed products rely on only two non-zero control observations.} As non-licensed products are over-represented in the treatment group, this underscores the importance of controlling for category composition in our main results. This also corresponds to higher market entry in licensed products, which limits follow-on use \citep{jackson_unlicensed_2009, milgrom_case_2011}.

We report all results with category controls, and note that all unconditional means presented in tables do not control for this category composition.

The main results show insignificant linear time trends, both on the main and extended samples. Table \ref{apx:tab:time} presents results with month fixed effects and quadratic time controls. None of the time controls show significant coefficients and nor do these controls change the main result significantly. Another concern is the selection of thresholds in the sample definition, especially given the high proportion of control products from March 1998. Table \ref{apx:tab:time_restricted} shows results with two alternative sample definitions, April to September 1998 and September 1997 to October 1998. Both show results consistent with the main results (the former relies only on two non-zero observationss in the control group).

To further rule out that our results are just a result of random patterns in time and across categories, we compare our main result to a set of results based on 1,000 random samples, stratified by category and application month \citep{young_channeling_2019}. The main coefficient is higher than 99 percent of these bootstrapped values.

Two placebo tests examine if the results could be due to unobserved product or firm differences. In Table \ref{apx:tab:placebo_product}, we alter the outcome variable to $\text{BackwardUse}_i$, the use of a frequency prior to focal product's introduction. As by definition, our main specification has only zero values on this alternative outcome, we expand the sample to products with prior use. We obtain a result indistinguishable from null (P .92), in support of the assumption that the treatment did not favor products with more attractive frequency combinations.

In Table \ref{apx:tab:placebo_firm}, we alter the treated products within firms. We take products from 1993--1997 (1992--1996 for extended sample) with new frequency combinations of each firm in the control and treatment groups. We assign placebo treatment to all products of firms that had at least one product truly treated. We obtain a result indistinguishable from null (P .69), in support of the assumption that the treatment did not favor products of firms that had higher follow-on use prior to the treatment period.

Together, these results present robust evidence of a significant increase of follow-on use of frequency combinations introduced in products whose exhibits were scanned and published by the government. We next examine which mechanisms drive these results.

\section{Mechanisms and extensions}
\label{sec:extensions}

To understand the mechanisms driving the increase in follow-on use in terms of firm type, we examine the timing, geography and firm and product characteristics. We rely on this extended sample in much of this analysis to avoid weakening the statistical power of the estimates either by splitting the sample or the outcome variable.

\begin{figure}[t]
    \centering
    \includegraphics[width=0.8\textwidth]{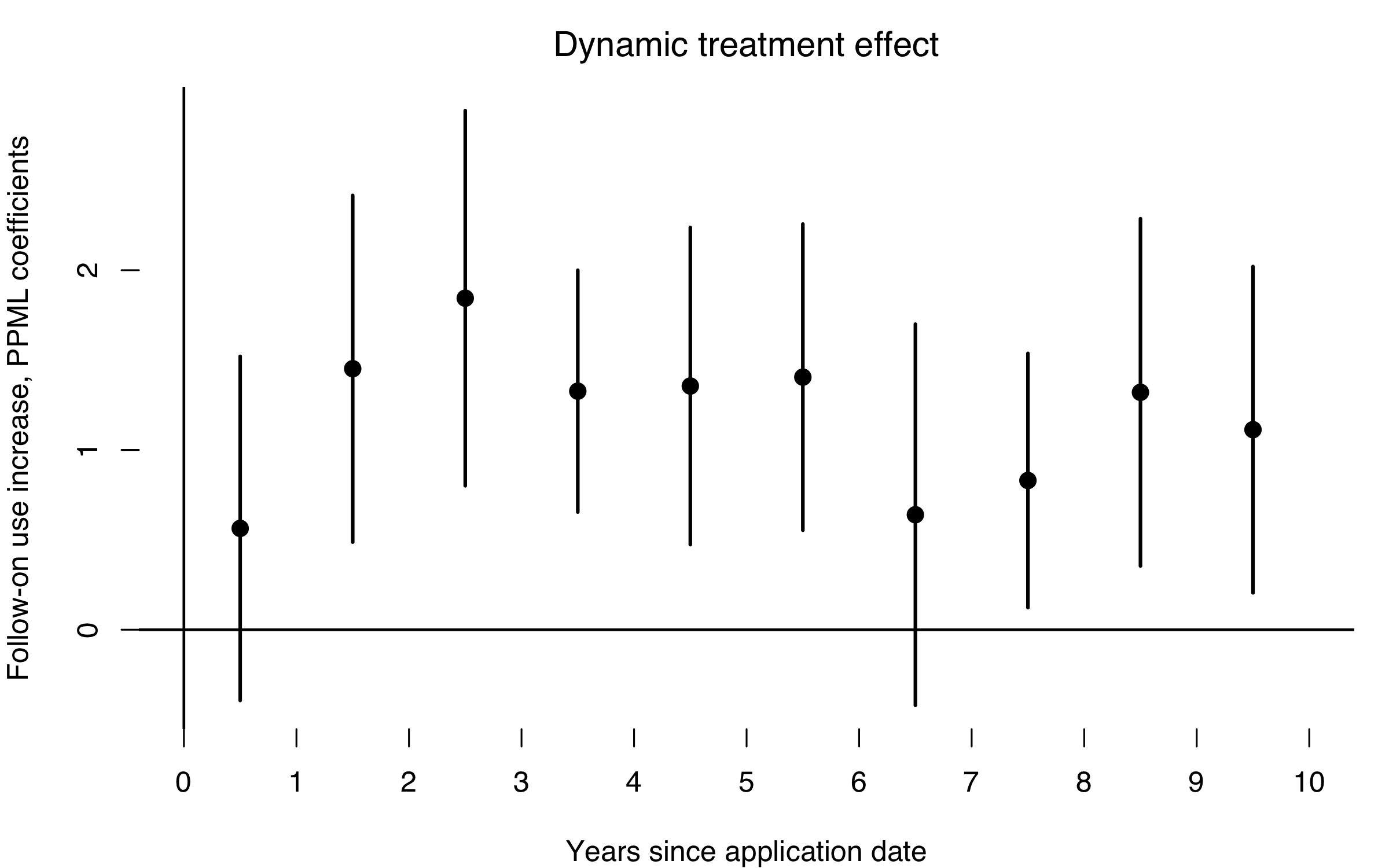}
    \caption{ForwardUse in time since application date.}
    \label{fig:dynamic}
    \footnotesize Coefficients $\beta_{\tau}$ ($\tau$ since product introduction date) from the specification
    \begin{math} 
        \log \mathbb{E}\left[\text{Apps}_{i\tau}\right]
        =
        \sum_{\tau} \beta_{\tau}
        \left(
        \text{GovernmentExhibits}_{i} \times \mathbf{1}\{\text{FollowYear}_{i\tau} = \tau\}
        \right)
        + \rho \text{ApplicationDate}_{i}
        + \lambda_{\tau}
        + \mathbf{X}_{i}^{\prime}\theta
        + \delta_{c(i)}
    \end{math}.
    Follows main results in Table \ref{tab:main} with extended control group (column 6). 95 percent confidence intervals, standard error clustered by applicant.

\end{figure}

We first examine the time dynamics and margin characteristics of the increase. Figure \ref{fig:dynamic} plots the treatment effect dynamically in time by interacting the treatment with years since the product application date. It shows an increasing effect in time, starting with a small and insignificant estimate within the first year of product introduction, and peaking in the third year at $1.83$. We interpret this as a lag in the absorption of the information available in the exhibits, and in the introduction of new products that use the same frequency combinations---comparable to that found in prior literature \citep{hall_nber_2001, hegde_patent_2023}. Surprisingly, the effect retains significance even after eight years. This may correspond to the continued use of the technologies in the first product, or the establishment of common technologies that are then routinely used in the industry---an entry of the information into common knowledge, or a technological lock-in.

To understand the margins of increase, we decompose our main effect into the probability of initial discovery (extensive margin) and the breadth of subsequent diffusion (intensive margin). The extensive margin analysis (columns 1 and 2) shows an approximate doubling of the number of follow-on products in the treatment groups, from a baseline of 13 percent. The intensive margin analysis shows a positive but insignificant increase. We note that because the estimate is based on post-treatment non-zero observationss, the estimate is principally downward-biased \citep{chen_logs_2024, lee_training_2009}. We further estimate covariate-adjusted Lee bounds in appendix \ref{apx:robustness} by estimating the increase among the always-taker population within quintiles of increased probability of treatment, yielding lower bound between 1.02 and upper bound 10.52, indicating a positive intensive margin effect \citep{lee_training_2009,semenova_generalized_2025}.

\begin{table}
    \input{tables/result_foreign.tex}

    \caption{Main result, country split}
    \label{tab:result_foreign}

    \footnotesize PPML estimates of the effect of government-scanned exhibits on follow-on innovation (\emph{ForwardUse}), split by U.S. and non-U.S. products. The sample includes products with new frequency combinations submitted between March 1997 and October 1998, excluding deregulated and self-exhibited products. Domestic (foreign) notes forward use by firms located in the same (different) country as the originator. Standard errors in parentheses, clustered by applicant.

\end{table}

We next examine the geography of the firms introducing the follow-on products. The period surrounding the policy shock saw significant trade liberalization, an increase in foreign imports to the U.S., and worries of foreign firms copying U.S. technology in these products. Many countries entered the World Trade Organization in 1995 and 1996, and China and Taiwan entered in 2001 and 2002. In this context, the transparency shock might have especially helped foreign firms adapt technology introduced in U.S. products.

Table \ref{tab:result_foreign} shows results split by originator and follow-on use country.\footnote{To achieve convergence and maintain sample size, we exclude \emph{Secrecy} control and only control for \emph{LicensedClass} in this specification.} The effect is clearly concentrated among foreign follow-on use of technologies introduced in products of U.S. firms (column 2), and still significant but smaller increase by U.S. domestic firms (column 1). The estimates on products introduced by non-U.S. firms are less definitive, but suggest little effect on domestic use of frequencies introduced in products by foreign firms (columns 3) and potentially some but insignificant increase in foreign use (column 4).

\begin{table}
    \input{tables/result_originator_entrant.tex}

    \caption{Main result, originator and entrant split}
    \label{tab:originator_entrant}

    \footnotesize PPML estimates of the effect of government-scanned exhibits on follow-on innovation (\emph{ForwardUse}), split by originator, competitor, incumbent and entrant products. The sample includes products with new frequency combinations submitted between March 1997 and October 1998, excluding deregulated and self-exhibited products. Standard errors in parentheses, clustered by applicant.

\end{table}

We next examine the characteristics of the follow-on products. Table \ref{apx:tab:simple_class} documents that the increase is concentrated among products in the same class as the original product (column 1). However, it also shows a marginally significant and lower increase of products in a different class, suggesting the technology contained in the original product had in some cases been adapted for uses in other product types. A split between simple (one frequency) and complex (multiple frequency) products also shows an increase in both, although only the latter being statistically significant. Overall, it shows that the increase was due to a wider technology follow-on use than just mere copies of simple devices.

Finally, we examine the market positions of firms introducing the follow-on products. One might not expect the originator to be directly affected by the increased availability of information about products they introduced to the market. Yet there might be a second-order effect through increased competitive pressure that might cause the originator to introduce more (or fewer) products \citep{aghion_effects_2009}.

Table \ref{tab:originator_entrant} splits the follow-on use by competitors and originators in columns 1 and 2. The follow-on use increase is clearly positive and significant among competitors (column 1), comparable in magnitude to the main pooled result. The increase in use among originators (column 2) is less definitive. Although similar in magnitude, the estimate is significantly less precise, which may be explained by a low baseline follow-on use. The estimate is insufficiently powered to support or rule out an increase of follow-on use by the originator. Table \ref{tab:originator_entrant} further splits follow-on use by incumbents and entrants, defined as second and later products introduced by the follow-on firm. The follow-on use increase is clearly positive and significant among incumbents (column 1), but smaller in magnitude and significance among new entrants.

Overall, the results document a clear increase in follow-on use among competitor and incumbent companies. More broadly, they are consistent with the transparency shock broadening the set of products being introduced to the market, including new uses of the technologies, and especially by foreign firms. We next examine evidence of the reaction of the originator to this increase.

\subsection{Originator response}

A natural question is if the transparency shock and associated competitive pressure had second-order effects on the originator firm. Schumpeterian models offer competing predictions for an incumbent facing an exogenously elevated threat of imitation. A firm might attempt to `escape competition' by accelerating radical innovation to re-establish a technological lead \citep{aghion_competition_2001}. To secure the returns on these new investments against further imitation, the originator might also substitute toward stronger formal innovation appropriation mechanisms such as patenting and secrecy \citep{cohen_protecting_2000, hall_choice_2014}.

The theory also allows for a `discouragement effect'. When lowered information frictions significantly reduce the expected rents of future innovations, the threat of rapid imitation can stifle future innovation \citep{aghion_effects_2009}. Rather than investing in the development of new technologies, the originator may instead respond defensively toward incremental innovation—rapidly exploiting their already-exposed technology to capture remaining market share before competitors fully saturate the space \citep{acemoglu_radical_2022}.

\begin{table}
    \centering
    \begin{adjustbox}{max width=\textwidth}
    \input{tables/result_originator_inclusive.tex}

    \end{adjustbox}
    \caption{Originator response results, inclusive}
    \label{tab:originator_inclusive}

    \footnotesize Estimates of the effect of government-scanned exhibits on response of the originator. The sample includes products with new frequency combinations submitted between March 1997 and October 1998, excluding deregulated and self-exhibited products. Standard errors in parentheses, clustered by applicant.

\end{table}

Table \ref{tab:originator_inclusive} presents estimates of the treatment effect on future behavior of the originator in the years following the introduction of the focal product. It shows no significant change in survival (column 1), a small but statistically inconclusive increase in product introductions and patenting (columns 2 and 5), and no significant change in patenting and products with secrecy and new frequency combination (columns 3 and 4).

Interpreting these findings requires caution, as firm-level outcomes are susceptible to within-firm treatment spillovers: firms in the control group might be exposed to treatment through other products and vice-versa, and further leakage may occur after the post-period when all new products were scanned.

To limit these spillovers, we exclude firms that had products in multiple treatment categories (no exhibits, self exhibits, government exhibits) during the treatment period, leaving 53 percent of firms (36 percent of products) in the sample. The results in Table \ref{apx:tab:originator_restricted} yield more negative but similarly inconclusive estimates. This restriction mechanically excludes firms with more products and changes the sample, creating a trade-off between sample restriction and spillovers. Overall, these limitations of the firm-level outcomes prevent us from inferring strong conclusions about the originator response.

\section{Conclusion}
\label{sec:conclusion}

Governments have long shaped the information environment in which firms innovate. Venetian medieval laws constrained the movement of glass workers to maintain a technological lead \citep{amato_island_1997}. The 18th-century British parliamentary rewards system actively encouraged innovation disclosure, paying the inventor Thomas Lombe for exhibiting silk working machines in the Tower of London \citep{burrell_parliamentary_2015}. The 1999 U.S. American Inventor's Protection Act accelerated the disclosure of patent applications, and likely increased technology diffusion \citep{hegde_patent_2023}. 

Today, debates over artificial intelligence regulations resemble those that were preoccupying policymakers at the beginning of wireless products regulations: their primary goal is to ensure the safety of an increasingly pervasive technology \citep{jackson_unlicensed_2009,greenstein_wireless_2026}. The disclosure and transparency requirements of such regulations would likely alter information frictions---as do government support for open access innovation, debates about algorithmic transparency, as well as recent closures of U.S. government websites \citep{gotfredsen_fighting_2025}. Intentionally or not, government regulations alter the information environmnent.

This paper documents the effect of one such government transparency policy---the launch of a public online database of detailed technical documentation for wireless products---on follow-on innovation. Exploiting the FCC's bulk-scanning of paper exhibits during the 1998 transition, we find that exposure raised the follow-on use of the frequencies in those products roughly four to seven times over the subsequent five years. The effect emerges with a lag, peaks around the third year after market entry, and persists for nearly a decade. It is driven by competitors and established firms rather than new entrants, and is largest when foreign firms build on technologies first introduced by U.S. firms, pointing to cross-border diffusion of frontier wireless technology. These magnitudes survive a wide set of time, category, and placebo checks, and the institutional record is consistent with scanning that did not target technologically promising applications.

We document large diffusion benefits from transparency, but we cannot measure the full counterfactual effect on the incentives to create the original technologies. The central balance between protecting the inventor and encouraging follow-on innovation remains, and our findings should not be read as implying that more disclosure is unambiguously desirable. Reassuringly, we find no evidence that exposed firms were harmed along observable margins, even as our firm-level estimates are empirically constrained. As modern economies produce and process increasing amounts of digital information, our results suggest that even incidental changes in information frictions can have large effects on private-sector innovation.

\newpage


\newpage

\bibliography{references.bib}

\end{document}


\setstretch{1.5} 

\maketitle

\tableofcontents

\newpage

\section{Institutional background}
\label{sec:background}

\subsection{FCC database as a source for reverse-engineering}
\label{sec:blogs}

\paragraph{`Using the FCC EAS for fun and profit'} \citet{baddeley_using_2016} explains how to navigate the FCC ID database to find external and internal photos, user manuals, and testing data before a product is released or disassembled. Excerpt:
\begin{quote}
`On any of [the FCC filings], we can look at the detailed information of the filing. This gives us all kinds of goodies, including the external and internal photos, user manual, and testing data. The photos are extremely important for finding motivation, as they will include pictures underneath shields for wireless modules. This lets you look at a product and determine exactly what is inside, which technologies they used, how they laid out the components, etc., and it's all free and easily available as soon as it's filed, which is usually before the product is even available for sale. Is there a product you like and you want to find out how they did it? Check the FCC database.'
\end{quote}

\paragraph{`Robbing the IoT Graveyard: A Foray into Info Gathering and Enumeration with the InternetVue 2100'} \citet{augusto_robbing_2021} exposes how querying the FCC database exposed full architectural block diagrams and an unpopulated JTAG pinout for a legacy IoT media receiver. Excerpt:
\begin{quote}
`As a new analyst [...], I was often told to check the FCC database before performing any hardware analysis, and for good reason. Every communications device needs to pass through rigorous FCC testing by an independent auditor, which will often expose things the company does not particularly want to see, including things like detailed schematics and communications protocols. [...] Using the FCC information, I was able to show that the two connections I had identified were likely JTAG and UART.'
\end{quote}

\paragraph{`A Crash Course in Hardware Hacking Methodology'} \citet{keys_crash_2024} details a hardware penetration testing methodology that relies on the FCC ID database for initial open-source intelligence (OSINT) and functional evaluation. Excerpt:
\begin{quote}
`Using the first device shown, navigating to fccid.io and entering the FCC ID […] we find that there was a request to make certain details confidential, but as shown below, we can still access images of the internal hardware and components along with technical information about wireless protocols in use. While these images may not seem like much at first glance, they hold fundamental details about internal components, debug ports and other specifications that would not otherwise be known unless the device was disassembled.'
\end{quote}

\paragraph{`Reverse engineering the Tempur-Pedic adjustable base remote control'} \citet{laplante_reverse_2019} describes the process of reverse engineering an adjustable base remote for smart home integration. Uses FCC database to extract technical details from test reports such as operating frequencies, channel mappings, and transmission patterns. Excerpt:
\begin{quote}
`We can lookup the FCC ID [...] to learn some useful information about the device. Confidentially requests by the manufacturer prevent us from accessing the BOM, schematic, and block diagram :/. In this case the most helpful document is the Test Report. [...] Page 13-14 detail that the remote transmits in pulses, e.g. on each key press. This implies that there is no persistent connection between the base and the remote, and that communication is probably unidirectional.'
\end{quote}

\paragraph{`Reverse Engineering the Maverick ET-732'} \citet{blake_reverse_2015} documents the extraction of transmission frequencies and modulation types from FCC verification documents to reverse engineer an RF barbecue thermometer. Excerpt:
\begin{quote}
`I also searched FCC filings using the Maverick's FCC ID of [...] which turned out to be extremely useful. In addition to verifying what I saw on the spectrum analyzer, I was able to download schematics and all sorts of other helpful information about the transmitter.'
\end{quote}

\paragraph{`Hardware Hacking - Dumping Flash Memory of a TrendNet-731BRv1 Router'} \citet{arch_cloud_labs_hardware_2024} outlines the process of using FCC database internal photos to visually identify and locate flash memory chips on a router's PCB for in-circuit firmware extraction. Excerpt:
\begin{quote}
`The [FCC documentation] among other things (there's a wealth of information here) tells me that my keyfob uses 2 distinct channels 433.66Mhz \& 434.18Mhz as its frequency of operation. That's good, it means my keyfob is more reliable, as it transmits on multiple channels. So we have the frequency of operation.'
\end{quote}

\paragraph{`Hideez Key 2 FAIL: How a good idea turns into a SPF (Security Product Failure)'} \citet{paganini_hideezkey_2021} shows how looking up an FCC ID revealed PCB layouts and specific microcontrollers as the first step in passive reconnaissance of a security token. Excerpt:
\begin{quote}
`First of all (even without attempting to open the token) we can immediately notice our best-hardware-hacking-friend: the FCC ID. […] This leads us to the first set of OSINT information regarding how the PCB looks like, what MCU is used and which frequency is in use by the DUT (Device Under Test). […] The Internal Photos PDF from the FCC database is particularly useful in case the DUT's PCB would have been buried under a hard layer of epoxy.'
\end{quote}


\subsection{Product examples}
\label{sec:examples}

\begin{figure}[ph]
    \centering
\includegraphics[width=\textwidth,keepaspectratio]{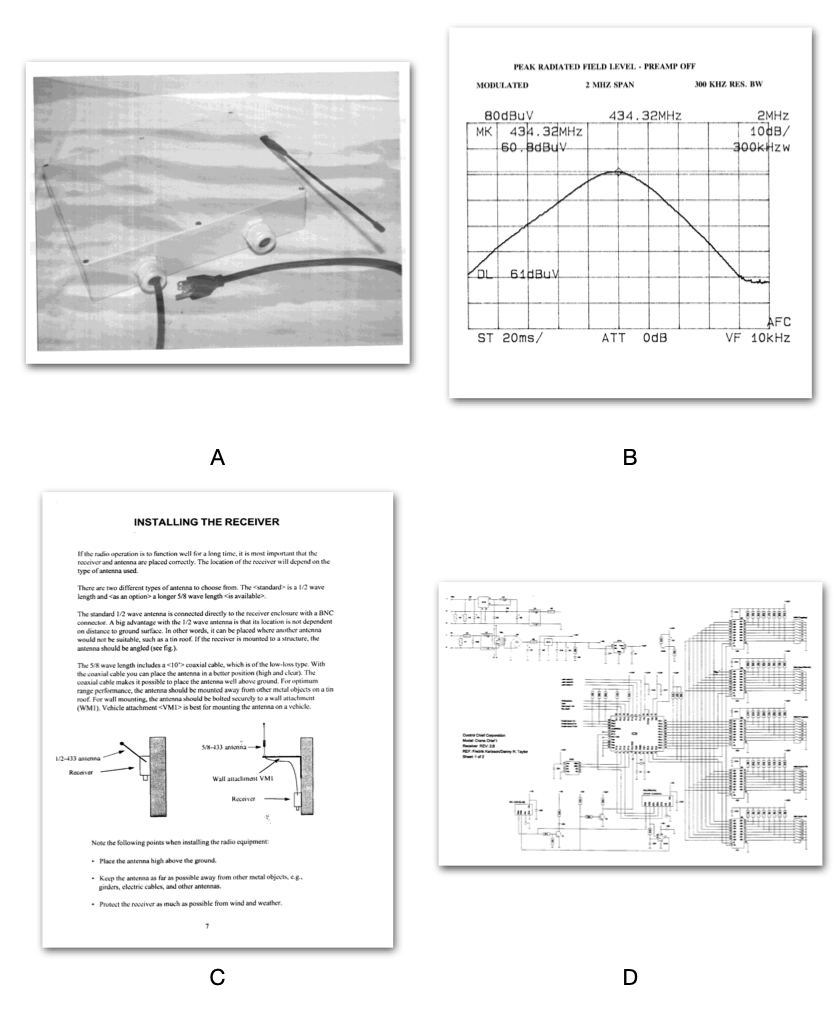}
    \caption{Product example, FCC ID CBFCRANET1}
    \footnotesize Excerpts from documentation submitted to the FCC as part of the equipment authorization process, and available online: A) external photos, B) test results, C) user manual excerpt, D) schematics.
    \label{fig:exhibit1}
\end{figure}

\begin{figure}[ph]
    \centering
\includegraphics[width=\textwidth,keepaspectratio]{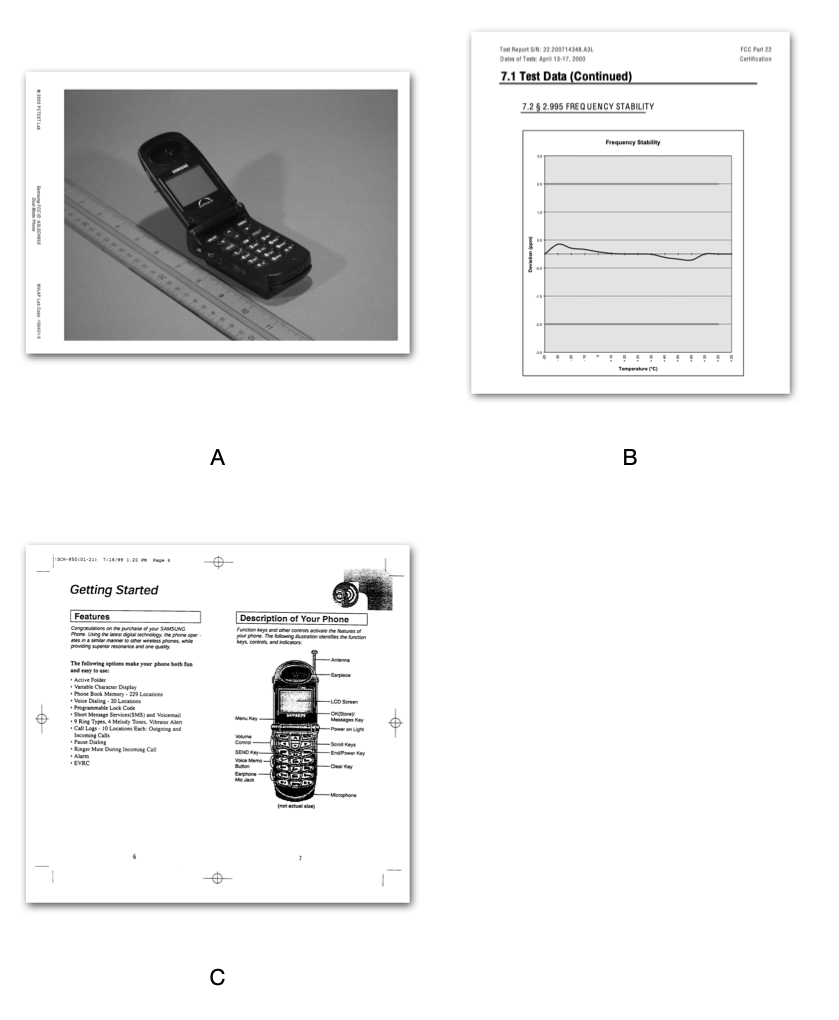}
    \caption{Product example, FCC ID A3LSCH855}
    \footnotesize Excerpts from documentation submitted to the FCC as part of the equipment authorization process, and available online: A) external photos, B) test results, C) user manual excerpt. Schematics were claimed confidential by the applicant.
    \label{fig:exhibit2}
\end{figure}

Two product examples illustrate the contents of the dataset. The product with FCC ID CBFCRANET1 is a crane control transmitter applied for by Control Chief Corporation on October 5, 1998. It operates on one frequency, 434--435 MHz, a previously unused range. Six other products used this frequency within the next five years, one receiver by the same company, and other products by different companies for similar use (industrial remote control transceivers and receivers, for example). Seven exhibits of this application are available online (see Figure \ref{fig:exhibit1} for excerpts). The firm did not claim secrecy on the exhibits, nor did it apply for a patent in the three years prior.

The product with FCC ID A3LSCH855 is a flip phone applied for by Samsung Electronics on July 18, 2000. It uses three frequencies, two at 824--849 MHz and one at 825--848 MHz. The combination of frequencies is not new, as two other firms (Sony Electronics and Sanyo Electric) used it in two different phones the previous year. This is the last product that used this frequency combination. 32 exhibits from this application are available online (see Figure \ref{fig:exhibit2} for excerpts); three additional exhibits are claimed confidential by Samsung: the block diagram, schematics, and circuit description. The firm applied for a patent in the three years prior.

\subsection{Validation of exhibit disclosure regimes}
\label{sec:exhibits}

Figure \ref{fig:exhibits_weekday} shows the distribution of products by disclosure regime and submission date day of the week. The ten governmnent exhibited products all have the same encoding sofware (`Acrobat Distiller 4.0 for Windows'), six have no author, and four share the same author (`VicodinES /CB /TNN'). This appears to be the signature of a computer hacker group active in late 1990s \citep{fairley_raney_new_1998}. In contrast, the ten self exhibited products have three different encoding softwares (albeit six have a dominant one, `Acrobat Distiller 4.0 for Windows'), two have no author, and eight have five different author values (including `jsoscia', `Microsoft Word', and one with `VicodinES /CB /TNN').

The metadata shows the vast majority of the self exhibits group consists of applicant-uploaded files, but we also occasionally see signs of government-scanned applications (here, the username 'jsoscia', a username of a contractor). These are cases of the exhibits being scanned after the application date, but before the grant date. As noted in the main text, we leave these products in the self exhibits group.

\begin{figure}[th]
    \centering
\includegraphics[width=\textwidth,keepaspectratio]{figures/figure_exhibits_weekday.png}
    \caption{Proportion of products by disclosure regime and submission day of the week}
    \footnotesize 1998 products, excluding deregulated products and products with no exhibits available.
    \label{fig:exhibits_weekday}
\end{figure}

Table \ref{tab:exhibits_metadata} shows metadata of the test report exhibit of twenty products.

\begin{table}[th]
    \centering
    \footnotesize
    \caption{Sample of exhibits metadata}
    \label{tab:exhibits_metadata}
    \resizebox{\textwidth}{!}{%
    \begin{tabular}{lllll}
    \toprule
    \textbf{FCCID} & \textbf{Exhibits} & \textbf{Format} & \textbf{Author} & \textbf{Software} \\
    \midrule
    LQP-R00 & Self & PDF & C:SCANDALLSCANDALL.EXE & Pixel Translations (PIXPDF Ver.1.38) \\
    N3E-10167 & Self & PDF & ADOBEPS4.DRV Version 4.24 & Acrobat Distiller 4.0 for Windows \\
    MGPCM-11 & Self & PDF & VicodinES /CB /TNN & Acrobat Distiller 4.0 for Windows \\
    B5KKRC16131 & Self & PDF & ADOBEPS4.DRV Version 4.24 & Acrobat Distiller 4.0 for Windows \\
    NNURHINSSL$*$ & Self & PDF & & Acrobat Distiller 4.0 for Windows \\
    CZ57RRKR & Self & PDF & Microsoft Word & Acrobat PDFWriter 3.02 for Windows \\
    NPR9810006000A & Self & PDF & & Acrobat Distiller 4.0 for Windows \\
    N5WNP1PSBSM01 & Self & PDF & Microsoft Word & Acrobat PDFWriter 3.02 for Windows \\
    NW6KAS-2000 & Self & PDF & jsoscia & Acrobat Distiller 4.0 for Windows \\
    NZENP-150 & Self & JPG & & \\
    \midrule
    CHP8BUHPB500-A & Government & PDF & VicodinES /CB /TNN & Acrobat Distiller 4.0 for Windows \\
    CMYKFK7935 & Government & PDF & & Acrobat Distiller 4.0 for Windows \\
    NZQ98C0001 & Government & PDF & & Acrobat Distiller 4.0 for Windows \\
    CJ6AT98-037 & Government & PDF & & Acrobat Distiller 4.0 for Windows \\
    HQXPC98010-30 & Government & PDF & & Acrobat Distiller 4.0 for Windows \\
    CLVJW-201HTRUDY & Government & PDF & VicodinES /CB /TNN & Acrobat Distiller 4.0 for Windows \\
    IW2PREMIOMTP6C & Government & PDF & & Acrobat Distiller 4.0 for Windows \\
    ICUVGA-GW806 & Government & PDF & & Acrobat Distiller 4.0 for Windows \\
    HAP91130 & Government & PDF & VicodinES /CB /TNN & Acrobat Distiller 4.0 for Windows \\
    MPO920-0123 & Government & PDF & VicodinES /CB /TNN & Acrobat Distiller 4.0 for Windows \\

    \bottomrule
    \end{tabular}%
    }
    \footnotesize Sample of products metadata of exhibits (test reports), ten of each exhibits availability regime. $*$ The product NNURHINSSL does not have test report available, the results are based on the PDF of external photos.

\end{table}

\section{Descriptives}
\label{sec:data}

\begin{figure}[p]
    \centering
    \begin{subfigure}[t]{0.49\textwidth}
        \centering
        \includegraphics[width=\textwidth]{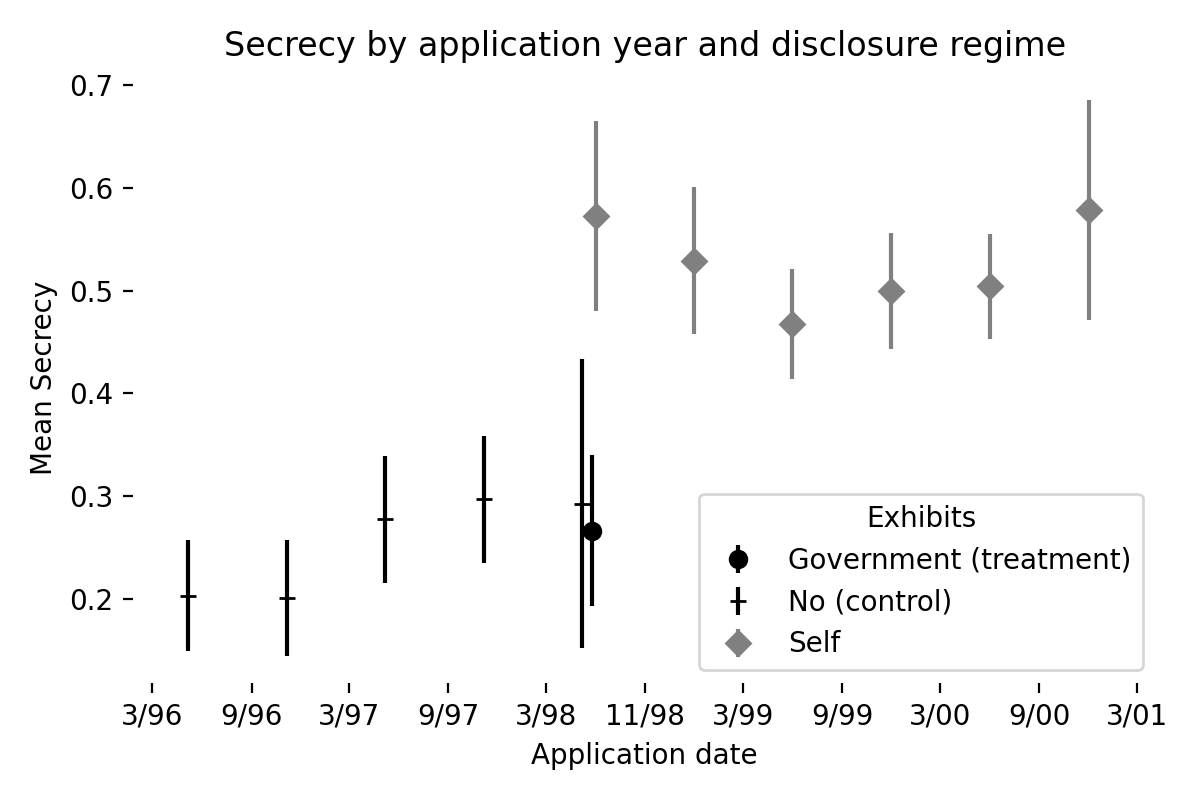}
        \caption{Secrecy by disclosure regime.}
    \end{subfigure}
    \hfill
    \begin{subfigure}[t]{0.49\textwidth}
        \centering
        \includegraphics[width=\textwidth]{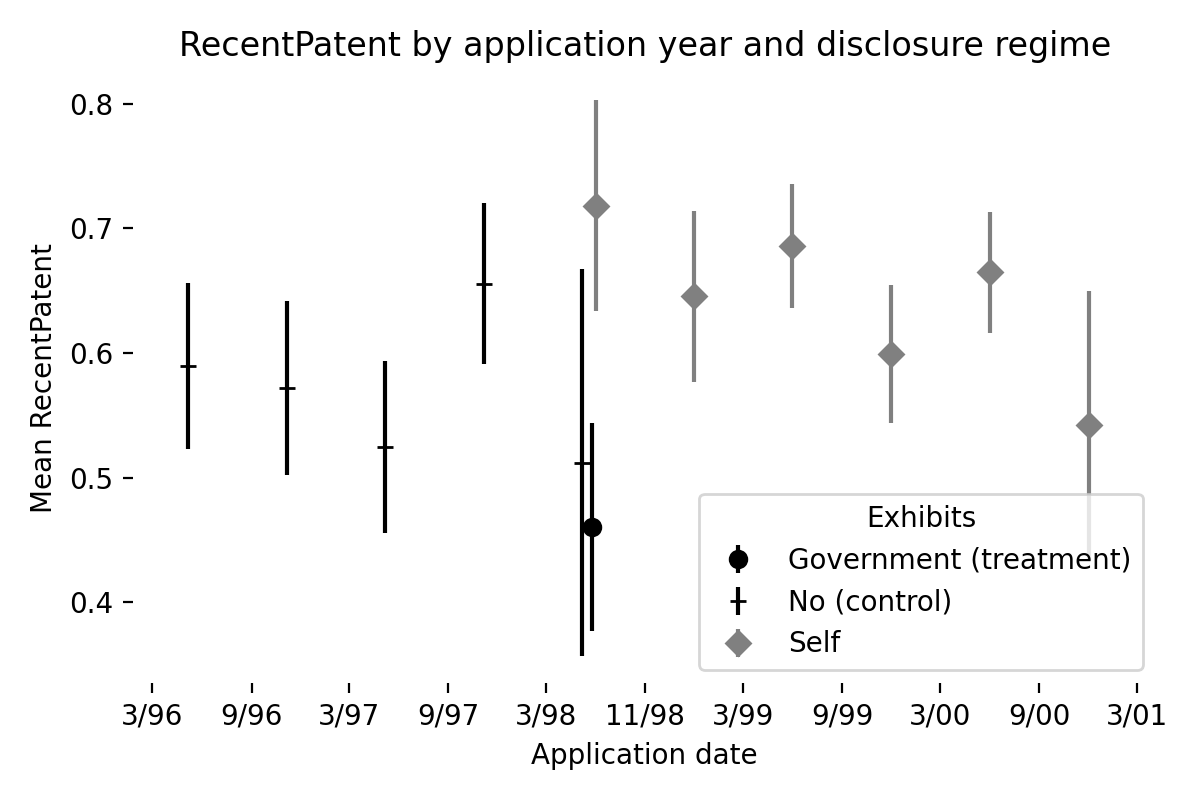}
        \caption{Patent by disclosure regime.}
    \end{subfigure}
    \caption{Bi-yearly means of secrecy and recent patenting around the transparency shock.}
    \footnotesize Same binning and sample restrictions as the bi-yearly plots in the main text.
    \label{fig:secrecy_patenting}
\end{figure}

Table \ref{tab:summary} shows summary statistics. Figure \ref{fig:secrecy_patenting} shows a trend in the increase in secrecy and patenting 10 years prior and post the relevant period.

\begin{table}[p]
\centering
\footnotesize
    \caption{Summary statistics}
    \resizebox{\textwidth}{!}{%
\begin{tabular}{lccccc@{\hspace{8mm}}ccccc@{\hspace{8mm}}ccccc}
\toprule
& \multicolumn{5}{c}{\textbf{All}} & \multicolumn{5}{c}{\textbf{Pre-period}}  & \multicolumn{5}{c}{\textbf{Post-period}}  \\
& Mean & SD & Q10 & Median & Q90 & Mean & SD & Q10 & Median & Q90 & Mean & SD & Q10 & Median & Q90 \\ \midrule 
\emph{A) Full sample} & &  &  &  &  & &  &  &  & \\
\hspace{3mm} NewFrequency & 0.3 & 0.46 & 0 & 0 & 1 & 0.28 & 0.45 & 0 & 0 & 1 & 0.32 & 0.47 & 0 & 0 & 1 \\
\addlinespace
\emph{B) NewFrequency = 1} & &  &  &  &  & &  &  &  & \\
\hspace{3mm} Secrecy & 0.54 & 0.5 & 0 & 1 & 1 & 0.18 & 0.38 & 0 & 0 & 1 & 0.7 & 0.46 & 0 & 1 & 1 \\
\hspace{3mm} RecentPatent & 0.62 & 0.48 & 0 & 1 & 1 & 0.56 & 0.5 & 0 & 1 & 1 & 0.65 & 0.48 & 0 & 1 & 1 \\
\hspace{3mm} FuturePatent & 0.64 & 0.48 & 0 & 1 & 1 & 0.62 & 0.49 & 0 & 1 & 1 & 0.65 & 0.48 & 0 & 1 & 1 \\
\hspace{3mm} NewEntrant & 0.13 & 0.33 & 0 & 0 & 1 & 0.12 & 0.33 & 0 & 0 & 1 & 0.13 & 0.33 & 0 & 0 & 1 \\
\hspace{3mm} Applicant U.S. & 0.66 & 0.48 & 0 & 1 & 1 & 0.76 & 0.43 & 0 & 1 & 1 & 0.61 & 0.49 & 0 & 1 & 1 \\
\hspace{3mm} Changed & 0.1 & 0.31 & 0 & 0 & 1 & 0.03 & 0.18 & 0 & 0 & 0 & 0.14 & 0.35 & 0 & 0 & 1 \\
\hspace{3mm} NewFrequency & 1 & 0 & 1 & 1 & 1 & 1 & 0 & 1 & 1 & 1 & 1 & 0 & 1 & 1 & 1 \\
\hspace{3mm} MultipleFilings & 0.15 & 0.36 & 0 & 0 & 1 & 0.06 & 0.24 & 0 & 0 & 0 & 0.19 & 0.4 & 0 & 0 & 1 \\
\hspace{3mm} LicensedClass & 0.53 & 0.5 & 0 & 1 & 1 & 0.72 & 0.45 & 0 & 1 & 1 & 0.45 & 0.5 & 0 & 0 & 1 \\
\hspace{3mm} Internet & 0.23 & 0.42 & 0 & 0 & 1 & 0.04 & 0.2 & 0 & 0 & 0 & 0.32 & 0.47 & 0 & 0 & 1 \\
\hspace{3mm} ForwardUse & 1.06 & 5.34 & 0 & 0 & 2 & 0.69 & 5.46 & 0 & 0 & 1 & 1.24 & 5.27 & 0 & 0 & 2 \\
\hspace{3mm} ForwardUse, Originator & 0.08 & 0.56 & 0 & 0 & 0 & 0.11 & 0.56 & 0 & 0 & 0 & 0.07 & 0.55 & 0 & 0 & 0 \\
\hspace{3mm} ForwardUse, Competitors & 0.98 & 5.21 & 0 & 0 & 1 & 0.57 & 5.25 & 0 & 0 & 1 & 1.17 & 5.18 & 0 & 0 & 2 \\
\bottomrule
\end{tabular}
    }
    \label{tab:summary}
    \footnotesize Summary statistics for main variables. Subsample of non-deregulated products within five years pre- and post- treatment, year 1998 excluded. P.1 and P.9 note the 10th and the 90th percentile values, respectively. 

\end{table} 

\subsection{Internet-capable products}
\label{sec:internet}

I code products connected to the internet based on U.S. 3G, 4G, 5G, and Wifi bands, which I match to applications based on frequencies, output, and year of introduction. Based on Wikipedia overviews and the website \url{https://www.rfwel.com}, Table \ref{tab:freqencies} summarizes the frequency thresholds I used in this coding. As 5G bands were established after the period of interest, I paid less attention to their encoding and did not encode 5G Wifi for the same reason. A product that has at least one frequency within the range, maximum output of more than 0.1 Watts (the upper limit for Class 1 Bluetooth), and year of introduction on or later than the introduction of the wireless band, is counted as having the relevant connectivity. 

\begin{table}[p]
\centering
\caption{U.S. internet bands}
\begin{tabular}{ccccc}
\toprule
\textbf{Technology}    & \textbf{Band} & \textbf{Uplink (MHz)} & \textbf{Downlink (MHz)} & \textbf{Year} \\
\midrule
Wifi                   & 2.4 GhZ       & \multicolumn{2}{c}{2412, 2462}                  & 1997                   \\
\midrule
\multirow{2}{*}{3G}    & PCS           & 1850, 1910            & 1930, 1990              & 1997                   \\
                       & UMTS          & 824, 849              & 869, 894                & 1997                   \\
\midrule
\multirow{19}{*}{4G}   & 2             & 1850 , 1910           & 1930 , 1990             & 2007                   \\
                       & 4             & 1710 , 1755           & 2110 , 2155             & 2007                   \\
                       & 5             & 824 , 849             & 869 , 894               & 2007                   \\
                       & 10            & 1710 , 1770           & 2110 , 2170             & 2007                   \\
                       & 12            & 699 , 716             & 729 , 746               & 2007                   \\
                       & 13            & 777 , 787             & 746 , 756               & 2007                   \\
                       & 14            & 788 , 798             & 758 , 768               & 2007                   \\
                       & 17            & 704 , 716             & 734 , 746               & 2007                   \\
                       & 23            & 2000 , 2020           & 2180 , 2200             & 2007                   \\
                       & 24            & 1626.5 , 1660.5       & 1525 , 1559             & 2007                   \\
                       & 25            & 1850 , 1915           & 1930 , 1995             & 2007                   \\
                       & 26            & 814 , 849             & 859 , 894               & 2007                   \\
                       & 27            & 807 , 824             & 852 , 869               & 2007                   \\
                       & 30            & 2305 , 23151          & 2350 , 2360             & 2007                   \\
                       & 66            & 1710 , 1780           & 2110 , 2200             & 2007                   \\
                       & 70            & 1695 , 1710           & 1995 , 2020             & 2007                   \\
                       & 71            & 663 , 698             & 617 , 652               & 2007                   \\
                       & 74            & 1427 , 1470           & 1475 , 1518             & 2007                   \\
                       & 85            & 698 , 716             & 728 , 746               & 2007                   \\
\midrule
\multirow{3}{*}{4G TDD} & 35            & \multicolumn{2}{c}{1850 , 1910}                 & 2007                   \\
                       & 36            & \multicolumn{2}{c}{1930 , 1990}                 & 2007                   \\
                       & 37            & \multicolumn{2}{c}{1910 , 1930}                 & 2007                   \\
\midrule
\multirow{8}{*}{5G}    & n2            & 1850 , 1910           & 1930 , 1990             & 2017                   \\
                       & n5            & 824 , 849             & 869 , 894               & 2017                   \\
                       & n12           & 699 , 716             & 729 , 746               & 2017                   \\
                       & n25           & 1850 , 1915           & 1930 , 1995             & 2017                   \\
                       & n66           & 1710 , 1780           & 2110 , 2200             & 2017                   \\
                       & n70           & 1695 , 1710           & 1995 , 2020             & 2017                   \\
                       & n71           & 663 , 698             & 617 , 652               & 2017                   \\
                       & n74           & 1427 , 1470           & 1475 , 1518             & 2017                   \\
\midrule
\multirow{3}{*}{5G TDD} & n41           & \multicolumn{2}{c}{2496 , 2690}                 & 2017                   \\
                       & n48           & \multicolumn{2}{c}{3550 , 3700}                 & 2017                   \\
                       & n77           & \multicolumn{2}{c}{3300 , 42002}                & 2017                   \\
\midrule
\multirow{2}{*}{5G mm}  & n260          & \multicolumn{2}{c}{37000 , 40000}               & 2017                   \\
                       & n261          & \multicolumn{2}{c}{27500 , 28350}               & 2017                  \\
\bottomrule
\end{tabular}
    \label{tab:freqencies}
    
\end{table}

\subsection{Deregulated products}
\label{sec:deregulated}

Correctly coding products deregulated around the same time period as the transparency shock is crucial for avoiding selection bias in the analysis. Products falling under different regulatory regimes are labeled by the applicable rule parts, which are noted on the FCC grant document, information from which is available in our data. I match those rule part codes in the data to changes outlined by the FCC.

Products under rule parts 2, 5, 15B, 17, 18, 21, 73, 74, 74E, 78, 87, 78 were deregulated from certification or notification procedure to declaration of conformity or verification \citep{federal_communications_commission_fcc_1995, federal_communications_commission_fcc_1997, federal_communications_commission_fcc_1998}. I therefore exclude all products with only these rule parts, including from applications prior to the policy change. Other changes, namely from notification to verification, and fee changes, are minor or administrative, and do not affect data availability. I therefore keep these observations in the dataset.

\begin{figure}[p]
    \centering
\includegraphics[width=\textwidth,height=\textheight,keepaspectratio]{figures/figure_deregulated.png}
    \caption{Deregulated products in time}
    \footnotesize Count of FCC product applications by year, split by deregulated and not deregulated. A decrease in the number of applications after 1998 is fully attributable to a decrease in deregulated product applications, which are excluded form the analysis. Grey area notes period from policy change announcement to end of uptake period.
    \label{fig:deregulation}
\end{figure}

I verify the accuracy of this approach in two ways. First, the proportion of deregulated products should decrease significantly after 1998, as applicants need to submit these products only voluntarily. Indeed, products in deregulated categories decreased from 79 percent in 1995 to 30 percent in 2000. Second, a decrease in applications during the time should be only in these product categories. This is indeed the case: Figure \ref{fig:deregulation} shows a drop in non-deregulated applications applications, but no drop among those that were not deregulated.

\subsection{Product category construction}
\label{sec:category}

We aggregate FCC product classes into broader economically meaningful categories. First, we group product classes into \emph{Licensed} (if at least one class mentions 'Licensed'), \emph{Part 15} (unlicensed, mentions 'Part 15'), and \emph{Other} (remaining products). In the rare case a product is in both in a Licensed and Part 15 class, the former takes precedent on the assumption that the use of a licensed spectrum affect the whole product. From these broad groups, we then separate individual class names with at least 5 percent of the products in the subset of data we use in our main specification. The result is seven distinct categories, summarized in Table \ref{tab:category}, which details their distribution.

\begin{table}[p]
\centering
\caption{Product category distribution}
\resizebox{\textwidth}{!}{%
    \begin{tabular}{llcc}
        \toprule
        Category & Classes & Count & Proporiton \\
        \midrule
        Licensed Non-Broadcast Station Transmitter & Licensed Non-Broadcast Station Transmitter & 60 & 0.23 \\
        Part 15 Spread Spectrum Transmitter & Part 15 Spread Spectrum Transmitter & 53 & 0.20 \\
        Part 15 Low Power Communication Device Transmitter & Part 15 Low Power Communication Device Transmitter & 27 & 0.10 \\
        Part 15 Low Power Transceiver, Rx Verified & Part 15 Low Power Transceiver, Rx Verified & 19 & 0.07 \\
        \addlinespace
        Licensed (other) & PCS Licensed Transmitter & 10 & 0.04 \\
        Licensed (other) & Licensed Non-Broadcast Transmitter Held to Ear & 5 & 0.02 \\
        Licensed (other) & Licensed Broadcast Station Transmitter & 3 & 0.01 \\
        Licensed (other) & Licensed Non-Broadcast Transmitter Worn on Body & 3 & 0.01 \\
        Licensed (other) & PCS Licensed Transmitter held to ear & 3 & 0.01 \\
        Licensed (other) & Communications Rcvr for use w/ licensed Tx and CBs | Licensed Non-Broadcast Station Transmitter & 2 & 0.01 \\
        Licensed (other) & Licensed Non-Broadcast Transmitter Held to Face & 2 & 0.01 \\
        Licensed (other) & Communications Rcvr for use w/ licensed Tx and CBs | Licensed Non-Broadcast Transmitter Held to Face & 1 & 0.00 \\
        Licensed (other) & Licensed Non-Broadcast Station Transmitter | Licensed Non-Broadcast Transmitter Held to Face & 1 & 0.00 \\
        Licensed (other) & Licensed Non-Broadcast Station Transmitter | Part 15 Low Power Communication Device Transmitter & 1 & 0.00 \\
        \addlinespace
        Part 15 (other) & Part 15 Security/Remote Control Transmitter & 10 & 0.04 \\
        Part 15 (other) & Part 15 Field Disturbance Sensor & 8 & 0.03 \\
        Part 15 (other) & Part 15 Class B Computing Device Peripheral | Part 15 Low Power Transmitter Below 1705 kHz & 4 & 0.02 \\
        Part 15 (other) & Part 15 Class B Computing Device Peripheral | Part 15 Spread Spectrum Transmitter & 3 & 0.01 \\
        Part 15 (other) & Part 15 Low Power Transceiver, Rx Certified | Superregenerative Receiver & 3 & 0.01 \\
        Part 15 (other) & Part 15 Low Power Transmitter Below 1705 kHz & 3 & 0.01 \\
        Part 15 (other) & Part 15 Class B Computing Device Peripheral | Part 15 Low Power Communication Device Transmitter & 2 & 0.01 \\
        Part 15 (other) & Part 15 Cordless Telephone System & 2 & 0.01 \\
        Part 15 (other) & Part 15 Anti-Pilferage Device & 1 & 0.00 \\
        Part 15 (other) & Part 15 Auditory Assistance Device (Transmitter) & 1 & 0.00 \\
        Part 15 (other) & Part 15 Class B Computing Device Peripheral | Part 15 Low Power Transceiver, Rx Verified & 1 & 0.00 \\
        Part 15 (other) & Part 15 Class B Computing Device/Personal Computer | Part 15 Spread Spectrum Transmitter & 1 & 0.00 \\
        Part 15 (other) & Part 15 Cordless Telephone Base Transceiver & 1 & 0.00 \\
        Part 15 (other) & Part 15 Low Power Communication Device Transmitter | Part 15 TV Interface Device & 1 & 0.00 \\
        Part 15 (other) & Part 15 Low Power Communication Device Transmitter | Part 18 Consumer Device & 1 & 0.00 \\
        Part 15 (other) & Part 15 Remote Control/Security Device Transceiver & 1 & 0.00 \\
        Part 15 (other) & Part 15 Security/Remote Control Transmitter | Superregenerative Receiver & 1 & 0.00 \\
        \addlinespace
        Other & Amplifier & 11 & 0.04 \\
        Other & Part 95 Family Radio Face Held Transmitter & 9 & 0.03 \\
        Other & Part 90 Location \& Monitoring Transmitter & 3 & 0.01 \\
        Other & Marine Radar & 1 & 0.00 \\
        Other & Part 80 HF Transmitter (GMDSS) & 1 & 0.00 \\
        Other & Part 80 MF Transmitter (GMDSS) & 1 & 0.00 \\
        Other & Part 95 Family Radio Body Worn Transmitter & 1 & 0.00 \\
        Other & Part 95 Family Radio Ear Held Transmitter & 1 & 0.00 \\
        \bottomrule
    \end{tabular}}

    \label{tab:category}
    \footnotesize Distribution of categories and product classes in the main dataset. 1998 products with new frequency combinations, excluding deregulate and self exhibited. Multiple classes separated by `|' sign.

\end{table}

\subsection{Extended covariates balance}
\label{sec:balance}

Table \ref{tab:balance} shows covariates balance with category fixed effects (A) and on an extended sample that includes products with previously used frequency combinations (B).

\begin{table}

    \caption{Covariate balance, extended.}
    \resizebox{\textwidth}{!}{%
    \centering
    \begin{tabular}[t]{lccc@{\hspace{8mm}}ccc@{\hspace{8mm}}ccc}

        \toprule
         & No & Government & Self & $\delta_1$ & SE($\delta_1$) & P($\delta_1$) & $\delta_2$ & SE($\delta_2$) & P($\delta_2$)\\
         &  & Exhibits &  & & & & & & \\
        & (1) & (2) & (3) & (4) & (5) & (6) & (7) & (8) & (9) \\
        \midrule
        \emph{(A) Category controls} &&&&&&&&& \\

        RecentPatent & 0.52 & 0.48 & 0.70 & -0.02 & 0.10 & 0.82 & 0.18 & 0.10 & 0.07\\
        Secrecy & 0.30 & 0.27 & 0.51 & 0.00 & 0.09 & 0.97 & 0.19 & 0.09 & 0.04\\
        ApplicantUS & 0.68 & 0.61 & 0.70 & -0.05 & 0.10 & 0.62 & -0.01 & 0.09 & 0.89\\
        NewEntrant & 0.09 & 0.11 & 0.18 & 0.00 & 0.05 & 0.98 & 0.12 & 0.06 & 0.05\\
        FrequencySimple & 0.45 & 0.60 & 0.57 & 0.03 & 0.09 & 0.76 & 0.06 & 0.09 & 0.54\\
        Internet & 0.16 & 0.17 & 0.31 & 0.06 & 0.06 & 0.35 & 0.14 & 0.07 & 0.04\\
        Changed & 0.09 & 0.10 & 0.14 & 0.02 & 0.05 & 0.75 & 0.05 & 0.06 & 0.37\\
        N & 44 & 149 & 182 &  &  &  &  &  & \\

        \addlinespace
        \multicolumn{10}{l}{\emph{(B) Extended sample: existing frequency combinations}} \\
   
        RecentPatent & 0.44 & 0.46 & 0.57 & 0.05 & 0.07 & 0.50 & 0.15 & 0.07 & 0.04\\
        Secrecy & 0.20 & 0.19 & 0.40 & 0.01 & 0.05 & 0.89 & 0.17 & 0.05 & 0.00\\
        ApplicantUS & 0.58 & 0.57 & 0.59 & 0.00 & 0.06 & 0.97 & -0.02 & 0.06 & 0.69\\
        NewEntrant & 0.12 & 0.12 & 0.19 & -0.02 & 0.03 & 0.47 & 0.07 & 0.04 & 0.08\\
        FrequencySimple & 0.68 & 0.73 & 0.68 & -0.08 & 0.06 & 0.20 & -0.10 & 0.07 & 0.17\\
        Internet & 0.08 & 0.07 & 0.18 & 0.05 & 0.03 & 0.09 & 0.11 & 0.03 & 0.00\\
        Changed & 0.05 & 0.06 & 0.12 & 0.02 & 0.03 & 0.34 & 0.08 & 0.04 & 0.02\\
        N & 153 & 442 & 429 &  &  &  &  &  & \\

        \addlinespace
        \multicolumn{10}{l}{\emph{(C) Extended sample: products since March 1997}} \\

        RecentPatent & 0.58 & 0.48 & 0.70 & -0.12 & 0.08 & 0.12 & 0.05 & 0.08 & 0.48\\
        Secrecy & 0.29 & 0.27 & 0.51 & -0.02 & 0.06 & 0.72 & 0.19 & 0.07 & 0.01\\
        ApplicantUS & 0.72 & 0.62 & 0.69 & -0.09 & 0.07 & 0.21 & -0.05 & 0.07 & 0.49\\
        NewEntrant & 0.13 & 0.11 & 0.18 & -0.03 & 0.04 & 0.43 & 0.07 & 0.05 & 0.11\\
        FrequencySimple & 0.44 & 0.60 & 0.57 & 0.07 & 0.07 & 0.30 & 0.10 & 0.07 & 0.16\\
        Internet & 0.20 & 0.17 & 0.30 & 0.04 & 0.05 & 0.39 & 0.11 & 0.06 & 0.04\\
        Changed & 0.08 & 0.10 & 0.14 & 0.01 & 0.05 & 0.79 & 0.05 & 0.05 & 0.35\\
        N & 455 & 151 & 185 &  &  &  &  &  & \\

        \bottomrule
        \end{tabular}

    }

    \footnotesize
    
    Extended covariates balance test. (1) mean value of not scanned products, (2) mean value of products scanned by the FCC or its contractor, (3) mean value of products self-scanned by the applicant, (4--6) and (7--9) the time-trend adjusted coefficient, standard error, and p-value $\delta$ from raw $X_i = \alpha + \delta \text{Exhibits}_i + \rho \text{ApplicationDate}_i + \gamma_c + \varepsilon_i$, where $X_i$ is the variable in the first column, $\text{Exhibits}_i$ is $1$ for government ($\delta_1$) and self-scanned ($\delta_2$), respectively, and $\gamma_c$ denotes category fixed effects. Panel A) shows results for the main sample, panel B) extends to all products between March and October 1998 excluding deregulated products, panel C) extends to products with new freqyency combinations sice 1997 (excluding deregulated products). Standard error clustered by Applicant.
    \label{tab:balance}

\end{table}

Table \ref{tab:treatment_category} estimates the relationship between selection into treatment and product categories ($\delta_c(i)$). Both products in the generic licensed category and in the `LicensedNon-BroadcastStationTransmitter' are significantly less likely to have exhibits scanned. The other categories are statistically comparable.

\begin{equation} \label{eq:treatment_cat}
    \text{GovernmentExhibits}_{i} = \delta_c(i) + \rho \text{ApplicationDate}_{i}  + \varepsilon_{i} 
\end{equation}

\begin{table}
    \caption{Treatment probability and categories}
    \begingroup
    \centering
    \begin{tabular}{lcc}
    \toprule
        & \multicolumn{2}{c}{Government exhibits}\\
                                                                & (1)      & (2)\\  
    \midrule 
    Category $=$ Licensed                                     & -0.2921  & -0.1778\\   
                                                                & (0.1112) & (0.0845)\\   
    Category $=$ LicensedNon-BroadcastStationTransmitter      & -0.2917  & -0.2273\\   
                                                                & (0.0865) & (0.0902)\\   
    Category $=$ Other                                        & -0.1753  & -0.0158\\   
                                                                & (0.1211) & (0.0778)\\   
    Category $=$ Part15LowPowerCommunicationDeviceTransmitter & -0.1764  & 0.0139\\   
                                                                & (0.1196) & (0.0585)\\   
    Category $=$ Part15LowPowerTransceiver,RxVerified         & 0.0041   & 0.0410\\   
                                                                & (0.0627) & (0.0609)\\   
    Category $=$ Part15SpreadSpectrumTransmitter              & -0.0976  & 0.0120\\   
                                                                & (0.0819) & (0.0632)\\   
        \\
    Application date                                          & Yes   & Yes\\   
        \\
    Observations                                              & 193      & 595\\  
    R$^2$                                                     & 0.13037  & 0.24431\\  
    Adjusted R$^2$                                            & 0.09747  & 0.23530\\  
    AIC                                                       & 201.46   & 552.92\\  
    BIC                                                       & 227.56   & 588.03\\  
    \bottomrule
    \end{tabular}
    \label{tab:treatment_category}
    \footnotesize 

    Coefficient estimates of the relationship between product categories and selection into treatment. Column (1) includes the main treatment sample, column (2) extends to all products with existing frequency combination. Standard errors clustered by applicant.

    \par\endgroup
\end{table}

\subsection{Covariates in time}
\label{sec:biyearly}

\begin{figure}[t]
    \centering
    \begin{subfigure}[t]{0.49\textwidth}
        \centering
        \includegraphics[width=\textwidth]{figures/figure_yearly_new_entrant.png}
        \caption{New entrant by disclosure regime.}
    \end{subfigure}
    \hfill
    \begin{subfigure}[t]{0.49\textwidth}
        \centering
        \includegraphics[width=\textwidth]{figures/figure_yearly_internet.png}
        \caption{Internet by disclosure regime.}
    \end{subfigure}
    \begin{subfigure}[t]{0.49\textwidth}
        \centering
        \includegraphics[width=\textwidth]{figures/figure_yearly_applicant_us.png}
        \caption{U.S. applicant by disclosure regime.}
    \end{subfigure}
    \hfill
    \begin{subfigure}[t]{0.49\textwidth}
        \centering
        \includegraphics[width=\textwidth]{figures/figure_yearly_multiple_apps.png}
        \caption{Multiple filings by disclosure regime.}
    \end{subfigure}
    \caption{Bi-yearly means around the transparency shock.}
    \footnotesize 
    \label{fig:biyearly_plots}
\end{figure}

\section{Robustness and extensions}
\label{robustness}

\begin{table}
    \input{tables/result_licensed.tex}
    \caption{Main results, licensed product frequencies split}
    \label{tab:licensed}

        \footnotesize PPML estimates of the effect of government-scanned exhibits on follow-on innovation (\\emph{ForwardUse}), licensed products split. The main sample includes products with new frequency combinations submitted between March and October 1998, excluding deregulated and self-exhibited products. Extended control sample includes observations since January 1997. Standard errors in parentheses, clustered by applicant.

\end{table}

\begin{table}
    \input{tables/result_time.tex}
    \caption{Main results, extended time controls}
    \label{tab:time}

        \footnotesize PPML estimates of the effect of government-scanned exhibits on follow-on innovation (\\emph{ForwardUse}), quadratic time and month fixed-effects. The main sample includes products with new frequency combinations submitted between March and October 1998, excluding deregulated and self-exhibited products. Extended control sample includes observations since March 1997. Standard errors in parentheses, clustered by applicant.

\end{table}

\begin{table}
    \input{tables/result_time_restricted.tex}
    \caption{Main results, time periods variations}
    \label{tab:time_restricted}

        \footnotesize PPML estimates of the effect of government-scanned exhibits on follow-on innovation (\\emph{ForwardUse}), varied time periods. The main sample includes products with new frequency combinations submitted between March and October 1998, excluding deregulated and self-exhibited products. Extended control sample includes observations since March 1997. Extended control sample includes observations between September 1997 and October 1998. Standard errors in parentheses, clustered by applicant.

\end{table}

Table \ref{tab:licensed} shows results split by licensed and non-licensed product classes. Table \ref{tab:time} shows results split by licensed and non-licensed product classes. Table \ref{tab:time_restricted} shows results with alternative time periods: April and September 1998 and September 1997 and October 1998.

\begin{table}
    \input{tables/result_placebo_product.tex}
    \caption{Product-level placebo}
    \label{tab:placebo_product}

        \footnotesize PPML estimates of the effect of government-scanned exhibits on prior frequency use (\\emph{BackwardUse}). The main sample includes products submitted between March and October 1998, excluding deregulated and self-exhibited products. Extended control sample includes observations since March 1997. Standard errors in parentheses, clustered by applicant.

\end{table}

\begin{table}
    \input{tables/result_placebo_firm.tex}
    \caption{Firm-level placebo}
    \label{tab:placebo_firm}

        \footnotesize PPML estimates of the firm-level placebo effect of government-scanned exhibits on follow-on innovation (\\emph{ForwardUse}). The main sample includes products in the main treatment and control groups introduced between 1993--1997. The main sample includes products in the main treatment and control groups introduced between 1991--1996. Standard errors in parentheses, clustered by applicant.

\end{table}

Table \ref{tab:placebo_product} presents product-level placebo, where we substitute the main specification outcome variable, $\text{ForwardUse}_i$, for $\text{BackwardUse}_i$. This counts the number of product that used the frequency combination prior to the focal product. As by definition, our main sample has only zero values on this alternative outcome, we expand the sample to all products.

\begin{table}
    \input{tables/result_extensive_intensive.tex}
    \caption{Intensive and extensive margin}
    \label{tab:extensive_intensive}

    \footnotesize PPML estimates of the effect of government-scanned exhibits on follow-on innovation (\emph{ForwardUse}), extensive and intensive margins. The sample includes products with new frequency combinations submitted between March 1997 and October 1998, excluding deregulated and self-exhibited products. Domestic (foreign) notes forward use by firms located in the same (different) country as the originator. Standard errors in parentheses, clustered by applicant.

\end{table}

Table \ref{tab:extensive_intensive} estimates the effect on the extensive (columns 1 and 2) and intensive margins (columns 3 and 4).

To account for the fact that treatment assignment is not strictly randomized but depends on observable characteristics (notably category and time), we implement a covariate-adjusted version of the Lee \citep{lee_training_2009} bounds proposed by \cite{semenova_generalized_2025}. Let $D \in \{0,1\}$ denote the treatment, $S \in \{0,1\}$ the selection indicator ( $\text{ForwardUse} > 0$), and $Y$ the intensive margin outcome. We are interested in bounding the average treatment effect for the always-takers subpopulation, $\Delta = E[Y(1) - Y(0) \mid S(1)=1, S(0)=1]$, under the standard monotonicity assumption that $S(1) \ge S(0)$.

Because standard Lee bounds may be biased when treatment depends on covariates $X$, we employ a propensity score subclassification approach. First, we estimate a baseline selection score, $p(X) = P(S=1 \mid X)$, using a fixed-effects logit model that includes all baseline controls and fixed effects. We then stratify the sample into five quintiles $g \in \{1, \dots, 5\}$ based on the predicted score $\hat{p}(X)$. 

Within each quintile $g$, we assume treatment is locally unconfounded and calculate the quintile-specific trimming fraction:
$$ q_g = \frac{P(S=1 \mid D=1, g) - P(S=1 \mid D=0, g)}{P(S=1 \mid D=1, g)} $$
Assuming $q_g > 0$ (i.e., treatment monotonically increases selection), we trim the upper (lower) $q_g$ fraction of the observed outcome distribution for the treated, selected units in quintile $g$ to compute the quintile-specific lower (upper) bound, $\Delta_g^L$ ($\Delta_g^U$). Finally, we aggregate these bounds to obtain the overall covariate-adjusted bounds:
$$ \Delta^B = \sum_{g=1}^5 w_g \Delta_g^B \quad \text{for } B \in \{L, U\} $$
where the weights $w_g$ are proportional to the number of selected control observations in quintile $g$. This weighting scheme consistently estimates the relative size of the always-selected population across quintiles.

\begin{table}
    \input{tables/result_simple_class.tex}
    \caption{Main result, class and simple split}
    \label{tab:simple_class}

    \footnotesize PPML estimates of the effect of government-scanned exhibits on follow-on innovation (\emph{ForwardUse}), split by class and product complexity. The sample includes products with new frequency combinations submitted between March 1997 and October 1998, excluding deregulated and self-exhibited products. Standard errors in parentheses, clustered by applicant.

\end{table}

Table \ref{tab:simple_class} extimates the effects by in-class and outclass follow-on products, and simple and complex original products.

\section{Originator}

\begin{table}
    \centering
    \begin{adjustbox}{max width=\textwidth}
    \input{tables/result_originator_main.tex}
    \end{adjustbox}
    \caption{Originator response results, restricted sample of firms}
    \label{tab:originator_restricted}

    \footnotesize Estimates of the effect of government-scanned exhibits on response of the originator. The sample includes products with new frequency combinations submitted between March 1997 and October 1998, excluding deregulated and self-exhibited products, and products of firms with both unambiguous treatment during the period. Standard errors in parentheses, clustered by applicant.

\end{table}

Table \ref{tab:originator_restricted} shows results of originator, exclusing products from firms with ambiguous treatment.

\newpage

\bibliography{references.bib}

%% file: preamble.tex
\usepackage[utf8]{inputenc}
\usepackage[margin=1in]{geometry}
\usepackage{pdfpages} 

\usepackage[hyphens]{xurl}
\usepackage[hidelinks,breaklinks]{hyperref} 
\usepackage{xr-hyper}

\usepackage{natbib} 

\usepackage{
    enumitem, 
    setspace, 
    multicol, 
    multirow, 
    graphics, 
    amsmath, 
    amssymb, 
    booktabs, 
    siunitx, 
    adjustbox, 
    float, 
}

\usepackage{titlesec}

\newcolumntype{d}{S[input-symbols = ()]}

\usepackage{subcaption}

\hypersetup{colorlinks = false,
linkbordercolor= white, citebordercolor= white, urlbordercolor= yellow,
,pdfborderstyle={/S/U/W 1},pdfborder=0 0 1}

\usepackage{placeins} 

%% file: tables/result_main.tex
\begingroup
\centering
\begin{tabular}{lcccccc}
   \toprule
   Sample & \multicolumn{3}{c}{Main} & \multicolumn{3}{c}{Extended control} \\ \cmidrule(lr){2-4} \cmidrule(lr){5-7}
    & \multicolumn{6}{c}{ForwardUse}\\
                          & (1)     & (2)     & (3)     & (4)      & (5)     & (6)\\  
   \midrule 
   GovernmentExhibits     & 2.27    & 1.77    & 1.92    & 1.83     & 1.28    & 1.30\\   
                          & (0.649) & (0.636) & (0.746) & (0.416)  & (0.381) & (0.331)\\   
                          &         &         &         &          &         &  \\  
   Linear Time Trend      & Yes     & Yes     & Yes     & Yes      & Yes     & Yes\\  
    \\
   Product Controls       & Yes     & Yes     & Yes     & Yes      & Yes     & Yes\\  
   Firm Controls          & Yes     & Yes     & Yes     & Yes      & Yes     & Yes\\  
   LicensedClass FE       &         & Yes     &         &          & Yes     & \\  
   Category FE            &         &         & Yes     &          &         & Yes\\  
    \\
   Observations           & 193     & 193     & 190     & 606      & 606     & 606\\  
   N Control              & 44      & 44      & 44      & 455      & 455     & 455\\  
   N non-zero (treatment) & 45      & 45      & 45      & 47       & 47      & 47\\  
   N non-zero (control)   & 5       & 5       & 5       & 58       & 58      & 58\\  
   Treatment mean         & 2.3     & 2.3     & 2.3     & 2.2      & 2.2     & 2.2\\  
   Control mean           & 0.20    & 0.20    & 0.20    & 0.26     & 0.26    & 0.26\\  
   Log-Likelihood         & -697.6  & -603.5  & -564.5  & -1,033.8 & -923.4  & -880.7\\  
   Pseudo R$^2$           & 0.13    & 0.25    & 0.29    & 0.21     & 0.30    & 0.33\\  
   AIC                    & 1,411.3 & 1,225.0 & 1,154.9 & 2,083.6  & 1,864.8 & 1,789.4\\  
   BIC                    & 1,437.4 & 1,254.3 & 1,197.1 & 2,118.9  & 1,904.5 & 1,851.1\\  
   \bottomrule
\end{tabular}
\par\endgroup

%% file: tables/result_foreign.tex
\begingroup
\centering
\begin{tabular}{lcccccc}
   \toprule
   Sample & \multicolumn{3}{c}{U.S.} & \multicolumn{3}{c}{Non-U.S.} \\ \cmidrule(lr){2-4} \cmidrule(lr){5-7}
   ForwardUse by          & All     & U.S.    & Non-U.S. & All     & U.S.   & Non-U.S. \\   
                          & (1)     & (2)     & (3)      & (4)     & (5)    & (6)\\  
   \midrule 
   GovernmentExhibits     & 1.38    & 0.955   & 2.23     & 1.35    & 1.71   & 1.15\\   
                          & (0.443) & (0.439) & (0.586)  & (0.905) & (1.44) & (0.912)\\   
                          &         &         &          &         &        &  \\  
   Linear Time Trend      & Yes     & Yes     & Yes      & Yes     & Yes    & Yes\\  
    \\
   Product Controls       & Yes     & Yes     & Yes      & Yes     & Yes    & Yes\\  
   Firm Controls          & Yes     & Yes     & Yes      & Yes     & Yes    & Yes\\  
   LicensedClass FE       & Yes     & Yes     & Yes      & Yes     & Yes    & Yes\\  
    \\
   Observations           & 422     & 422     & 422      & 184     & 184    & 184\\  
   N Control              & 329     & 329     & 329      & 126     & 126    & 126\\  
   N non-zero (treatment) & 27      & 20      & 22       & 20      & 9      & 16\\  
   N non-zero (control)   & 34      & 30      & 10       & 24      & 10     & 17\\  
   Treatment mean         & 2.6     & 1.4     & 1.2      & 1.7     & 0.90   & 0.78\\  
   Control mean           & 0.22    & 0.17    & 0.06     & 0.36    & 0.11   & 0.25\\  
   Log-Likelihood         & -657.3  & -442.0  & -293.1   & -232.3  & -130.3 & -152.9\\  
   Pseudo R$^2$           & 0.31    & 0.25    & 0.36     & 0.34    & 0.42   & 0.20\\  
   AIC                    & 1,328.6 & 898.0   & 600.2    & 478.7   & 274.5  & 319.8\\  
   BIC                    & 1,357.0 & 926.3   & 628.5    & 501.2   & 297.0  & 342.3\\  
   \bottomrule
\end{tabular}
\par\endgroup

%% file: tables/result_originator_entrant.tex
\begingroup
\centering
\begin{tabular}{lcccc}
   \toprule
   ForwardUse by          & Competitor & Originator & Incumbent & Entrant \\   
                          & (1)        & (2)        & (3)       & (4)\\  
   \midrule 
   GovernmentExhibits     & 1.28       & 1.25       & 1.37      & 0.951\\   
                          & (0.352)    & (0.810)    & (0.342)   & (0.442)\\   
                          &            &            &           &  \\  
   Linear Time Trend      & Yes        & Yes        & Yes       & Yes\\  
    \\
   Product Controls       & Yes        & Yes        & Yes       & Yes\\  
   Firm Controls          & Yes        & Yes        & Yes       & Yes\\  
   Category FE            & Yes        & Yes        & Yes       & Yes\\  
    \\
   Observations           & 606        & 606        & 606       & 606\\  
   N Control              & 455        & 455        & 455       & 455\\  
   N non-zero (treatment) & 38         & 14         & 46        & 19\\  
   N non-zero (control)   & 45         & 17         & 54        & 12\\  
   Treatment mean         & 1.9        & 0.25       & 1.9       & 0.34\\  
   Control mean           & 0.19       & 0.07       & 0.23      & 0.03\\  
   Log-Likelihood         & -764.3     & -211.7     & -776.7    & -165.0\\  
   Pseudo R$^2$           & 0.34       & 0.21       & 0.32      & 0.36\\  
   AIC                    & 1,556.6    & 451.3      & 1,581.5   & 358.1\\  
   BIC                    & 1,618.3    & 513.0      & 1,643.2   & 419.8\\  
   \bottomrule
\end{tabular}
\par\endgroup

%% file: tables/result_originator_inclusive.tex
\begingroup
\centering
\begin{tabular}{lccccc}
   \toprule
   Model & OLS & \multicolumn{4}{c}{PPML} \\ \cmidrule(lr){2-2} \cmidrule(lr){3-6}
                          & Survival & NProducts & NProductsNew & NProductsSecrecy & NProductsPatent\\  
                          & (1)      & (2)       & (3)          & (4)              & (5)\\  
   \midrule 
   GovernmentExhibits     & 0.012    & 0.206     & 0.058        & 0.062            & 0.160\\   
                          & (0.068)  & (0.210)   & (0.263)      & (0.268)          & (0.204)\\   
                          &          &           &              &                  &  \\  
   Linear Time Trend      & Yes      & Yes       & Yes          & Yes              & Yes\\  
   Product Controls       & Yes      & Yes       & Yes          & Yes              & Yes\\  
   Firm Controls          & Yes      & Yes       & Yes          & Yes              & Yes\\  
    \\
   Observations           & 606      & 606       & 606          & 606              & 606\\  
   N Control              & 455      & 455       & 455          & 455              & 455\\  
   N non-zero (treatment) & 111      & 97        & 87           & 80               & 64\\  
   N non-zero (control)   & 349      & 341       & 295          & 286              & 264\\  
   Treatment mean         & 0.74     & 11.8      & 5.3          & 5.8              & 8.7\\  
   Control mean           & 0.77     & 12.5      & 7.5          & 6.9              & 11.0\\  
   Log-Likelihood         & -313.6   & -6,268.9  & -3,467.6     & -3,496.9         & -6,081.7\\  
   Pseudo R$^2$           & 0.09     & 0.16      & 0.27         & 0.26             & 0.22\\  
   AIC                    & 655.2    & 12,565.8  & 6,963.2      & 7,021.9          & 12,191.4\\  
   BIC                    & 716.9    & 12,627.5  & 7,024.9      & 7,083.6          & 12,253.1\\  
   \bottomrule
\end{tabular}
\par\endgroup